\documentclass[journal]{IEEEtran}
\IEEEoverridecommandlockouts
\usepackage[english]{babel}
\usepackage{amsthm}
\usepackage{titlesec}
\usepackage[ruled,vlined]{algorithm2e}
\usepackage{algcompatible}
\usepackage{cite}
\usepackage{url}
\usepackage{hyperref}

\usepackage[pdftex]{graphicx}
\usepackage{subfigure}
\usepackage{float}
\usepackage{ragged2e}
\usepackage[labelfont=bf,font=small,justification=justified,singlelinecheck=false]{caption}
\usepackage{etoolbox}
\makeatletter
\patchcmd{\@makecaption}{\centering}{\justifying}{}{}
\makeatother
\usepackage{tikz}
\usepackage{array}
\usepackage{changepage}
\usepackage[utf8]{inputenc}
\usepackage{pgfplots} 
\usepackage{pgfgantt}
\usepackage{pdflscape}
 \usepackage{relsize}
\usepackage[export]{adjustbox}
\pgfplotsset{compat=newest} 
\pgfplotsset{plot coordinates/math parser=false}
\pgfplotsset{compat=1.18}
\usepackage{pgfplots}
\usetikzlibrary{spy}

\usepackage{cite}
\usepackage{amsmath,amssymb,amsfonts}
\usepackage{dsfont} 
\usepackage{stfloats}
\newtheorem{innercustomgeneric}{\customgenericname}
\providecommand{\customgenericname}{}
\newcommand{\newcustomtheorem}[2]{%
  \newenvironment{#1}[1]
  {%
   \renewcommand\customgenericname{#2}%
   \renewcommand\theinnercustomgeneric{##1}%
   \innercustomgeneric
  }
  {\endinnercustomgeneric}
}

\newcustomtheorem{customthm}{Theorem}
\newcustomtheorem{customlemma}{Lemma}
\newcustomtheorem{customprop}{Proposition}
\newcustomtheorem{customcor}{Corollary}
\usepackage{lipsum}
\usepackage{amsmath}
\usepackage[nolist,printonlyused]{acronym} 
\usepackage{amssymb}
\usepackage{mathtools}
\usepackage{url}
\usepackage{graphicx}  
\usepackage{float}  

\newcommand{\yy}{\mathbf{y}}
\newcommand{\xx}{\mathbf{x}}
\newcommand{\uu}{\mathbf{u}}
\newcommand{\bb}{\mathbf{b}}

\newcommand{\hh}{\mathbf{h}}
\newcommand{\nn}{\mathbf{n}}

\newcommand{\pp}{\mathbf{p}}

\newcommand{\XXi}{\boldsymbol{\Xi}}
\newcommand{\rr}{\mathbf{r}}
\newcommand{\ssb}{\mathbf{s}}

\newcommand{\rmr}{{\textrm{r}}}

\newcommand{\blkdiag}{\textrm{blkdiag}}

\newcommand{\Mbs}{{M_{\rm{BS}}}}
\newcommand{\Mue}{M_{\rm{UE}}}
\newcommand{\veccs}[1]{ {\rm{vec}}\big(#1\big)  }

\newcommand{\bs}{{\normalfont{\textrm{BS}}}}
\newcommand{\yyb}{\boldsymbol{y}}
\newcommand{\YYb}{\boldsymbol{Y}}
\newcommand{\nnb}{\boldsymbol{n}}
\newcommand{\NNb}{\boldsymbol{N}}
\newcommand{\ue}{{\normalfont{\textrm{UE}}}}

\newcommand{\rmd}{{\textrm{D}}}
\newcommand{\ZZ}{\mathbf{Z}}

\newcommand{\ptot}{{p_\textrm{tot}}}

\newcommand{\rmp}{{\textrm{p}}}

\newcommand{\BB}{\mathbf{B}}

\newcommand{\UU}{\mathbf{U}}

\newcommand{\review}[1]{{\color{black}{ #1}}}
\newcommand{\reviewAfter}[1]{{\color{black}{ #1}}}
\newcommand{\GG}{\mathbf{G}}
\newcommand{\YY}{\mathbf{Y}}

\newcommand{\HH}{\mathbf{H}}
\newcommand{\ff}{\mathbf{f}}
\newcommand{\VV}{\mathbf{V}}

\newcommand{\QQ}{\mathbf{Q}}
\newcommand{\FF}{\mathbf{F}}
\newcommand{\PP}{\mathbf{P}}
\newcommand{\XX}{\mathbf{X}}

\newcommand{\DD}{\mathbf{D}}
\newcommand{\CC}{\mathbf{C}}

\newcommand{\Tr}{\text{Tr}}
\newcommand{\diag}{\text{diag}}

\newcommand{\hermit}{\mathsf{H}}

\usepackage{accents}

\newcommand*{\ddt}[1]{%
	\accentset{\mbox{\large ..}}{#1}}

\newcommand{\TT}{\mathbf{T}}

\newcommand{\MM}{\mathbf{M}}

\newcommand{\atantwo}{{\rm{atan2}}}

\newcommand{\AAb}{\mathbf{A}}

\newcommand{\norm}[1]{\left\lVert#1\right\rVert}

\newcommand{\dd}{\mathbf{d}}

\newcommand{\vecc}{\text{vec}}
\newcommand{\pphi}{\boldsymbol{\phi}}
\newcommand{\EE}{\mathbf{E}}
\newcommand{\GGc}{\mathbf{G}}

\newcommand{\atx}{\mathbf{a}}

\newcommand{\tx}{{\textrm{Tx}}}
\newcommand{\rx}{{\textrm{Rx}}}
\newcommand{\cc}{\mathbf{c}}

\acrodef{SISO}[SISO]{single-input single-output}
\acrodef{AP}[AP]{access point}
\acrodef{AR}{autoregressive}
\acrodef{UE}[UE]{user equipment}
\acrodef{ULA}[ULA]{uniform linear array}
\acrodef{ML}[ML]{maximum likelihood}
\acrodef{CPU}[CPU]{central processing unit}
\acrodef{DL}{donwlink}
\acrodef{FPP}[FPP]{Feasible-point pursuit}
\acrodef{LoS}[LoS]{line of sight}
\acrodef{NLoS}[NLoS]{non-line-of-sight}
\acrodef{RCS}[RCS]{radar cross section}
\acrodef{AoD}[AoD]{angle of departure}
\acrodef{AoA}[AoA]{angle of arrival}
\acrodef{CRB}[CRB]{Cramer-Rao bound}
\acrodef{FIM}[FIM]{Fisher information matrix}
\acrodef{AN}[AN]{artificial noise}
\acrodef{RCP}[RCP]{residual clutter power}
\acrodef{SINR}[SINR]{signal-to-interference-plus-noise ratio}
\acrodef{SNR}[SNR]{signal-to-noise ratio}
\acrodef{QoS}[QoS]{quality of service}
\acrodef{SDR}[SDR]{semi-definite relaxation}
\acrodef{SDP}[SDP]{semi-definite program}
\acrodef{ISAC}[ISAC]{Integrated sensing and communications}
\acrodef{PLS}[PLS]{physical layer security}
\acrodef{SIC}[SIC]{successive interference cancellation}
\acrodef{CSI}[CSI]{channel state information}
\acrodef{CSIT}[CSIT]{channel state information at the transmitter}
\acrodef{CSIR}[CSIR]{channel state information at the receiver}
\acrodef{UL}[UL]{uplink}
\acrodef{DL}[DL]{downlink}
\acrodef{MSE}[MSE]{ mean square error}
\acrodef{TFR}[TFR]{target-to-floor ratio}
\acrodef{MUI}[MUI]{multi-user interference}
\acrodef{RIS}[RIS]{Reconfigurable intelligent surface}
\acrodef{AO}[AO]{alternating optimization}
\acrodef{SIMO}[SIMO]{Single Input Multiple Output}
\acrodef{MISO}[MISO]{multiple-intput single output}
\acrodef{MIMO}[MIMO]{multiple-input multiple-output}
\acrodef{MU}{multi-user}
\acrodef{BS}[BS]{base station}
\acrodef{CEE}[CEE]{channel estimation error}
\acrodef{CCP}[CCP]{convex-concave procedure}
\acrodef{MRT}[MRT]{ maximum-ratio transmission}
\acrodef{MRC}[MRC]{ maximum-ratio combining}
\acrodef{MM}[MM]{Minorization-Maximization}
\acrodef{PSD}[PSD]{positive semi-definite}
\acrodef{RZF}[RZF]{Regularized zero forcing}
\acrodef{CRZF}[CRZF]{Centralized regularized zero forcing}
\acrodef{RZF}[RZF]{regularized zero forcing}
\acrodef{LPZF}[LPZF]{Local Partial zero forcing}
\acrodef{LZF}[LZF]{Local zero forcing}
\acrodef{NF}[NF]{Near Field}
\acrodef{FF}[FF]{Far Field}
\acrodef{BD}[BD]{Beyond-diagonal}
\acrodef{OFDM}[OFDM]{orthogonal frequency division multiplexing}
\acrodef{MAPRT}[MAPRT]{maximum a-posteriori ratio test}
\acrodef{LRT}[LRT]{ likelihood ratio test}
\acrodef{CDF}[CDF]{Cumulative distribution function}
\acrodef{UPA}[UPA]{uniform planar array}
\acrodef{LMMSE}[LMMSE]{linear minimum mean square error}
\acrodef{MMSE}[MMSE]{minimum mean square error}
\acrodef{SE}[SE]{spectral efficiency}
\acrodef{CNR}[CNR]{clutter to noise ratio}
\acrodef{SCNR}[SCNR]{signal to clutter and noise ratio}
\acrodef{SOCP}[SOCP]{second order cone program}
\acrodef{TTD}[TTD]{true time delay}
\acrodef{PS}[PS]{phase shifter}
\acrodef{PEB}[PEB]{Position error bound}
\acrodef{JRC}[JRC]{Joint radar and communication}
\acrodef{GP}[GP]{gradient projection}
\acrodef{MO}[MO]{Manifold Optimization}
\acrodef{DEB}[DEB]{direction error bound}
\acrodef{REB}[REB]{range error bound}
\acrodef{LS}[LS]{least-squares}
\acrodef{MUSIC}[MUSIC]{MUltiple SIgnal Classification}
\acrodef{SR}[SR]{sum rate}
\acrodef{WMMSE}[WMMSE]{weighted-minimum mean square error}
\acrodef{BCD}[BCD]{block-coordinate descent}
\acrodef{SOCP}[SOCP]{second-order cone program}
\acrodef{DFT}[DFT]{discrete Fourier transform}
\acrodef{PSO}[PSO]{particle swarm optimization}
\acrodef{SIRV}[SIRV]{spherically-invariant random vector}
\acrodef{MF}[MF]{matched filter}
\acrodef{SCR}[SCR]{signal to clutter ratio}
\acrodef{CFAR}[CFAR]{constant false alarm rate}
\acrodef{RA}[RA]{range-angle}
\acrodef{RV}[RV]{range-velocity}
\acrodef{MU-MIMO}{multi-user multiple input multiple output}
\acrodef{NMSE}{normalized mean squared error}
\acrodef{SV}[SV]{singular value}
\acrodef{NS}[NS]{null-space}

\begin{document}

\title{On the Impact of Channel Aging and Doppler-Affected Clutter on OFDM ISAC Systems}

\author{
Steven~Rivetti,~\IEEEmembership{Student Member,~IEEE},
Gabor Fodor,~\IEEEmembership{Fellow,~IEEE}\\
Emil~Björnson,~\IEEEmembership{Fellow,~IEEE}, 
Mikael~Skoglund,~\IEEEmembership{Fellow,~IEEE}
  \thanks{
This work was supported by the SUCCESS project (FUS21-0026), funded by the Swedish Foundation for Strategic Research. G. Fodor was supported by the Swedish Strategic Research (SSF) FUS21-0004 SAICOM project.}
\thanks{The authors are with the School of Electrical Engineering and Computer Science (EECS), KTH Royal Institute of Technology, 11428 Stockholm, Sweden. This work was carried out while the first author was an intern at Ericsson Research, Stockholm, Sweden. Gabor Fodor is also with Ericsson Research, Stockholm, Sweden. } 

}

\maketitle

\begin{abstract}
The temporal evolution of the propagation environment plays a central role in integrated sensing and communication (ISAC) systems. A slow-time evolution manifests as channel aging in communication links, while a fast-time one is associated with structured clutter with non-zero Doppler. Nevertheless, the joint impact of these two phenomena on ISAC performance has been largely overlooked. This paper addresses this research gap in a network utilizing orthogonal frequency division multiplexing waveforms. Here, a base station simultaneously serves a user equipment (UE) device and performs monostatic sensing.
Channel aging is captured through an autoregressive model with Bessel correlation decay. In contrast, clutter is modeled as a collection of uncorrelated, coherent patches with non-zero Doppler, resulting in a Kronecker-separable covariance structure.
We propose an aging-aware channel estimator that uses prior pilot observations to estimate the time-varying UE channel, characterized by a non-isotropic multipath fading structure.
\reviewAfter{The clutter's structure enables a novel low-complexity clutter estimation and suppression procedure: clutter statistics are estimated from raw data and subsequently used to suppress the clutter's action, after which range–angle and range–velocity maps are computed.}
We evaluate the influence of frame length and pilot history on channel estimation accuracy and demonstrate substantial performance gains over block fading in low-to-moderate mobility regimes. 
\reviewAfter{
The proposed procedure is applied in a clutter-dominated environment, demonstrating that effective clutter suppression can be achieved under practical configurations.}We analyze the robustness of our proposed procedure against non-separable clutter by introducing a controllable degree of non-separability. \reviewAfter{Our results highlight the benefit of sensing streams and that the proposed clutter estimation and suppression procedure can withstand a moderate degree of non-separability. }
\end{abstract}
\section{introduction}

\IEEEPARstart{S}{ixth}-generation wireless networks are envisioned to support not only high-rate communications but also native sensing and environmental awareness. \ac{ISAC} has emerged as a key enabling paradigm in which sensing and communication tasks cooperate rather than competing for the ever more scarce spectral resources \cite{liu2022integrated}.
Among the many possible sensing topologies, monostatic sensing utilizing massive \ac{MIMO} at cellular \ac{BS} is of interest to the standardization and research communities due to its implementation advantages by utilizing large antenna arrays,
advanced signal processing capabilities, and the ability to make the communication signal readily available to the sensing receiver. In addition, cellular monostatic sensing does not require costly synchronization between multiple \acp{BS} \cite{Babu:24} or between \ac{UE} devices and \acp{BS}\cite{keskin2022optimal}. However, the monostatic operation sensing requires partitioning the antennas between transmission and reception, which differs from conventional communication-only deployments.
Furthermore, to facilitate task integration, an \ac{ISAC} system splits its available resources, such as transmit power and bandwidth, between the two tasks \cite{Fang:23}. The allocation of these resources is the primary factor determining the trade-off between sensing and communication. Examples of optimal resource allocation aimed at achieving the optimal trade-off can be found in \cite{ rivetti2024secure, AnLiu:22}.

A fundamental design choice in \ac{ISAC} systems is whether to pursue full integration between sensing and communication or to introduce some degree of separation between them. In this regard, \cite{Liu:25} provides a tutorial on how sensing can be performed with a waveform optimized for communications, thus achieving maximum integration.
However, this degree of integration introduces the so-called deterministic–random tradeoff: sensing benefits from deterministic signals while communication needs random ones. Introducing a degree of separation between the tasks can mitigate this tradeoff. In this regard, \cite{Baig:23} shows how space separation, achieved through the allocation of dedicated sensing streams in addition to the communication ones, achieves better \ac{AoA} estimation performance than a time separation scheme, where each task is carried out individually during different timeslots.
It is also worth mentioning that the geographical separation between a target and the communication \acp{UE} might make task separation unavoidable 
\cite{rivetti2024clutter,salman2024sensing}.
While sensing inherently relies on fast-time temporal variations of \ac{LoS} channels, communication channels, which are usually dominated by their \ac{NLoS} parts, show a slow-time temporal evolution caused by \ac{UE} mobility and environmental changes.
This temporal evolution is often simplified through the adoption of a block-fading model, which assumes a block-wise constant channel with every block being an independent realization. Although mathematically appealing, this model fails to capture channel aging \cite{Truong:13}, directly impacting the communication performance.
Channel aging can be modeled by accounting for statistical characteristics of the channel's temporal correlation \cite{Fodor:21,kong2015sum}.

Furthermore, recent advances in receiver design and channel estimation/prediction in communication systems indicate that pilot spacing (frame size) can help mitigate the negative effects of channel aging. This is a notoriously difficult problem to deal with: \cite{fodor2023optimizing} derives an upper bound on the \ac{SE} of a single-antenna \ac{UE}, finding the pilot spacing dictating the optimal trade-off between \ac{CSI} quality and \ac{SE}.
\cite{Daei:25} extends the previous analysis to non-stationary aging Rician channels, designing a multi-frame structure for data transmission, finding the optimal pilot spacing and power control strategy. \cite{Bjornson2015b} studies how hardware impairments affect the optimal pilot spacing.
A data-driven approach has been investigated in \cite{Yuan:20}, where the temporal correlation pattern is extracted by a convolutional neural network and the \ac{CSI} is predicted through an \ac{AR} network.

Somewhat surprisingly, the impact of channel aging on the communication performance of \ac{ISAC} systems in general and monostatic \ac{ISAC} systems in particular has not been fully investigated.
In \cite{chen2023ripoff}, a tracking-oriented approach is employed, where the aging time is defined as the number of blocks elapsed since the last pilot transmission, i.e., the number of blocks whose \ac{CSI} relies solely on prediction. \ac{CSI} is initially acquired through pilot transmission and \ac{MMSE} estimation, while 
prediction is achieved through a Kalman filter.
In spite of the fact that clutter is an unavoidable aspect affecting any radar system, many previous works fail to address the clutter's temporal evolution and its extended nature. In \cite{luo2024yolo}, a beam-squint-based ISAC scheme is proposed, aimed at near-field monostatic operation in the terahertz frequency band. Here, clutter is defined as a static grid of point-like reflectors. Although mathematically appealing, this does not capture the temporal evolution and spatial structure of realistic clutter environments. 

On the other hand, \cite{lu2025mimoclutt} allows for clutter to be composed of moving point-like scatterers, which nonetheless fails to account for the coherent, spatially extended structure of real clutter patches, which usually have non-negligible angular, Doppler, and delay spreads.
Clutter suppression is critical to the performance of \ac{ISAC} systems in terms of object detection and parameter estimation. For instance, \cite{Vinogradova:23} proposes an eigenvalue-based detection scheme applied to a narrow-band ISAC system, where the clutter space distribution is modeled as a set of patches and Toeplitz time-correlated noise.
The same target detection problem in the presence of patched clutter is tackled in \cite{demir2024ris}: here, the monostatic operation of a \ac{RIS} assisted ISAC network is analyzed, resulting in the proposal of a clutter distribution-agnostic target detector and a \ac{RIS} phase-shift optimization algorithm.

In this paper, we argue that ignoring or oversimplifying clutter leads to the design of less useful sensing pipelines in \ac{ISAC} systems.
In light of these recent advances in the design of \ac{ISAC} systems, we aim to analyze the impact of environmental changes on a monostatic ISAC network. The said network employs \ac{OFDM} waveforms to perform monostatic sensing in an urban scenario.
More specifically, our paper asks the following two important questions, which have not been conclusively answered by previous works:

\begin{itemize}
    \item \review{What is the impact of channel aging in an ISAC scenario where the \ac{BS} serves a multiple-antenna \ac{UE}}? How can one leverage the known channel temporal correlation structure to one's advantage?
    \item What is a realistic yet mathematically tractable model for clutter in an urban scenario ? How can one suppress the clutter action on the observed sensing data?
\end{itemize}

In light of these questions, the main contributions of our paper are the following:

\begin{itemize}
    \item We model the spatial and time correlation of the \ac{BS}-\ac{UE} channel. Its spatial covariance matrix is characterized in \ref{C_k_eq}; its temporal correlation follows a first-order \ac{AR} process with Bessel correlation decay.
    Lemma \ref{lemma hk} and its corollaries propose an aging-aware \ac{MMSE} channel estimator.
    \item Inspired by airborne and synthetic aperture radar techniques \cite{greenewald2016robust, ward1998space}, We propose a Kronecker-separable clutter covariance model, which is novel within the \ac{ISAC} environment.
    The clutter's covariance matrix is equal to the Kronecker product of the space, time, and frequency covariance matrices. Each of these matrices is defined in section \ref{cov mod sec}.
    \item \reviewAfter{A novel clutter estimation and suppression procedure is proposed. Starting from the raw data, clutter's second-order statistics are estimated, which are then used to suppress the former's action. Finally, \ac{RA} and \ac{RV} maps are computed to assess the achieved clutter suppression and the quality of the sensing representation available to downstream detection and estimation algorithms.}
    \item \reviewAfter{We assess the performance of the proposed clutter estimation and suppression procedure to non-separable clutter by introducing a controllable degree of non-separability. The adopted perturbation model allows us to simulate model mismatches in the eigenvalues and eigenvectors. }
    \item \reviewAfter{We numerically evaluate the performance of the proposed communication channel estimator and clutter estimation and suppression procedure in the considered single-UE setting. 
    The results show the advantages brought by leveraging the communication channel temporal correlation and how the clutter estimation-and-suppression procedure can withstand moderate levels of non-separability.    
    }
\end{itemize}

The remainder of this article is organized as follows: Section \ref{sys mod sec} shows the system model of the investigated monostatic \ac{OFDM} \ac{ISAC} network. Section \ref{cov mod sec} \review{provides the models for the UE's channel space and time covariances, a full characterization of the clutter second-order statistics, and the proposed non-separable robustness analysis }. Section \ref{ch est sec} presents the proposed aging-aware \ac{MMSE} channel estimator, giving a statistical characterization of its channel estimate and channel estimation error. Section 
\reviewAfter{
\ref{sec pipeline} presents the proposed estimation and suppression procedure, starting from the clutter covariance estimators in \ref{sec pipeline}-A.
Section \ref{sec prec} then shows the precoding techniques implemented in the article, consisting of MMSE precoders \cite{marzetta2016fundamentals} for the communication streams and null-space projected beamsweeping for the sensing stream. Finally, Section \ref{num res} showcases the performance of the system.}

\emph{Notation}: Boldface lowercase and uppercase letters denote vectors and matrices, respectively.
The trace of the matrix $\XX$ is denoted by $\Tr(\XX)$.
$\diag(\xx)$ represents the stacking of $\xx$ on the main diagonal of a matrix.
The creation of a tensor from a set of vectors is denoted by the following notation $\YY =\left\{[\yy_{1,c},\dots,\yy_{B,c}]\right\}_{\forall c}\in \mathds{C}^{A \times B \times C}$; conversely $\yy_{b,c}= \{ \YY\}_{b,c}$.
The notation $\mathcal{CN}(0,\sigma^2)$ represents the circularly symmetric complex Gaussian distribution with variance $\sigma^2$. 
Given the matrix $\XX \in \mathds{C}^{A \times B}$, the notation $\YY=\veccs{\XX}\in \mathds{C}^{AB}$ stacks the columns of said matrix to form a vector. The same applies to a tensor $\XX \in \mathds{C}^{A \times B \times C}$: $\yy=\veccs{\XX}\in \mathds{C}^{ABC}$.
Given the vector $\xx$, the notation $\mathcal{T}(\xx)$ denotes the Toeplitz matrix built from $\xx$. 
Given the tensor $\YY \in \mathds{C}^{A \times B \times C}$, the notation $\XX=\mathcal{U}_1(\YY)$ denotes the unfolding of $\YY$ along the first dimension, resulting in the matrix $\XX \in  \mathds{C}^{A \times B C} $. On the other hand, given the same matrix $\XX$, mentioned above, $\YY=\mathcal{F}_1(\XX)$ denotes the folding operation, retrieving the original tensor. \review{Given the angular interval $\Delta=[\theta,\vartheta]$, we denote by $|\Delta|$ its length, i.e. $|\Delta| = \vartheta-\theta$  }

\vspace{-3mm}
\section{System Model}\label{sys mod sec}

\begin{figure}[t!]
\begin{center}
   \resizebox{0.43\textwidth}{!}{
     \includegraphics[]{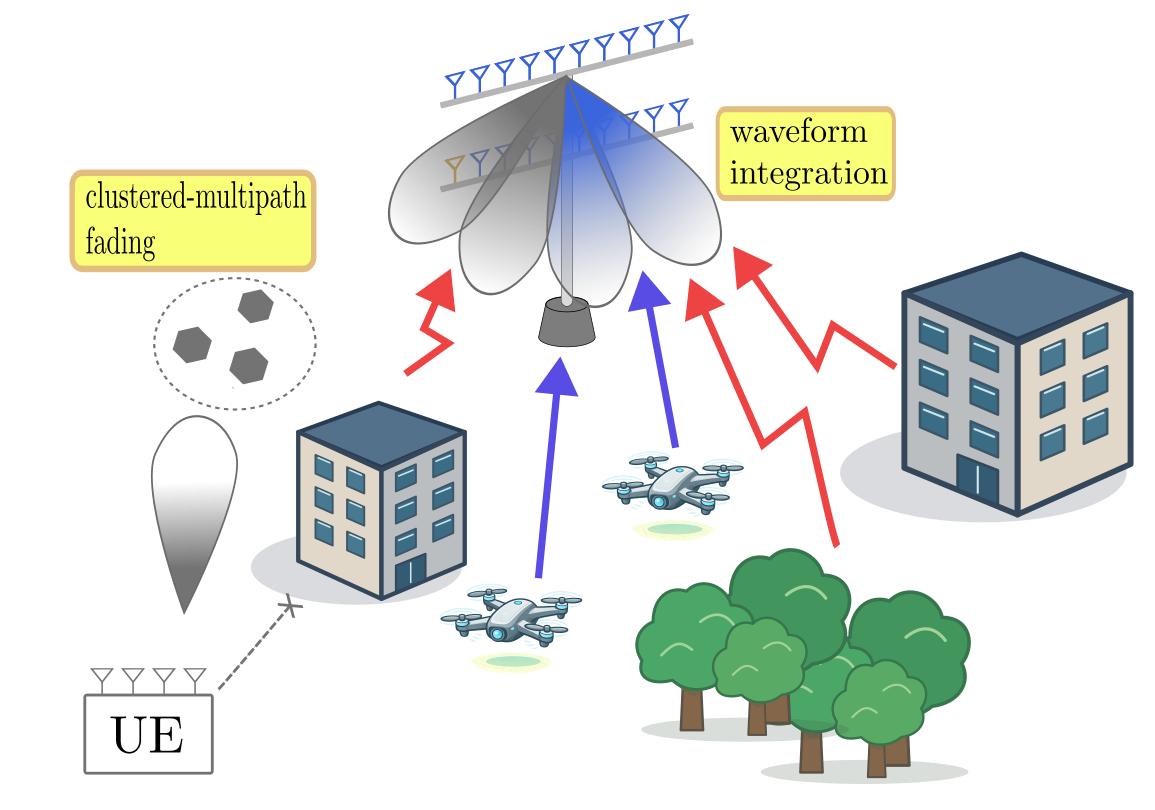}}
      \vspace{-2mm}
	 \caption{A single-cell network, where the \ac{BS} uses multi-antenna transmission to provide \ac{ISAC} services. This downlink signal is used to transmit pilots for channel estimation
    \cite{daei2024improved} and as a radar probe for monostatic sensing \cite{liyanaarachchi2023joint}. 
    The blue arrows represent target echoes, whereas the red arrows represent ground clutter echoes.}
    \label{scenario}
\end{center}
  \vspace{-4mm}
\end{figure}
 
We consider the single-cell scenario shown in Figure \ref{scenario}, where a
\ac{BS}, equipped with $\Mbs$ receiving and transmitting antennas, 
serves one communication \ac{UE} device, equipped with $\Mue$ antennas.
The \ac{DL} channels between the \ac{BS} and the \ac{UE} are estimated by transmitting \ac{DL} pilot signals and assuming instantaneous feedback by the \ac{UE}
\footnote{\review{
This assumption is introduced to enable estimation of \ac{CSIT}. The communication side of this work focuses on channel estimation quality, which is agnostic to any UE feedback mechanism.
}}.
\reviewAfter{At the same time, the transmitted waveform is used as a radar probing signal to collect monostatic echoes from $L$ targets.}
The system implements \ac{OFDM} transmission around carrier frequency $f_c$, over $V$ active subcarriers with a subcarrier spacing of $\Delta_f$. 
Here $T=1/\Delta_f + T_\textrm{cp} $ is the symbol duration, with $T_\textrm{cp}$ being the cyclic prefix duration. 
The communication channel between the \ac{BS} and the \ac{UE}  at time $t$ is denoted by $\HH_{t} \in \mathds{C}^{\Mue \times \Mbs}$. 
Due to the assumed frequency-domain block-fading structure, we omit the subcarrier index from $\HH $: the modeling considerations in the sequel refer to an arbitrary coherence block.
\begin{figure}[t!]
\begin{center}
   \resizebox{0.45\textwidth}{!}{
    \input{images/pilot_table/pilot_table_img}}
	  \caption{Representation of the pilot and data symbols transmission structure in the time-frequency domain. This transmission occurs on the communication channel, which is estimated at the start of each pilot slot. }
       \label{pilot table}
		\end{center}
  \vspace{-6mm}
\end{figure}
Under these assumptions,
let us build the vectorized channel of the \ac{UE}, denoted by 
$\hh_{t}=\vecc(\HH_{t})\in \mathbb{C}^{\Mbs\Mue}$, 
by stacking the columns of $\HH_{t}$. 
The vectorized channel follows a correlated Rayleigh fading distribution, where the covariance matrix follows a Kronecker model \cite{bengtsson2006some}: this model is a good fit since the UE and BS are 
in the respective far fields and the multipath is originated by a set 
of communication clusters
\footnote{\review{The Kronecker correlation model gives a tractable mathematical covariance formulation. We reserve the study of non-separable correlation models, such as Weichselberger models, \cite{debbah2022uplink} for future work. 
}}.
A communication cluster (referred to as ``cluster" in the sequel) is defined as a set of closely spaced scatterers in the proximity of the UE, generating small-scale fading \cite{38901}.
Therefore, $\hh_{t} \sim \mathcal{CN}(\mathbf{0},\CC)$, where 
\begin{align}
\CC=\CC_{\bs}\otimes\CC_{\ue } \in \mathds{C}^{\Mbs\Mue \times \Mbs\Mue}.
\end{align}
Here, $\CC_{\bs} \in \mathds{C}^{\Mbs \times \Mbs}$, 
and $\CC_{\ue} \in \mathds{C}^{\Mue \times \Mue}$ are the BS and UE side correlation matrices respectively.
We further assume that $\HH_{t}$ follows an aging structure in time, a model that aims to leverage the channel's correlation between timeslots, a degree of freedom not exploited when a block fading channel structure is assumed \cite{truong2013effects}.
We model this time aging as a first-order \ac{AR} process. Since each spatial component undergoes the same channel aging, the AR process state transition matrix is a scaled identity matrix \cite{fodor2023optimizing, abeida2010data}, where the scaling component controls the correlation decay over time.
Under these assumptions,the correlation between $\hh_{t}$ and $\hh_{t+iT}$ is defined as$\mathds{E}[\hh_{t}\hh^\hermit_{t+iT}]=\CC_k\zeta(iT),$ 
where $\zeta(iT)$ captures the correlation decay in time slot $i$.
Without loss of generality and for the sake of clarity, this notation assumes that the time unit is $T$ and therefore we will write $iT = i$ in the sequel.
Let us collect the channel realizations of $\Delta$ time slots in the vector 
$\hh_\Delta  \in \mathds{C}^{\Mbs\Mue\Delta}$.  This is a zero-mean complex Gaussian vector 
whose covariance matrix is defined as 
\begin{align}
    \mathds{E}[\hh_\Delta\hh_\Delta^\hermit]=\CC 
    \otimes \mathcal{T}\Big([\zeta(0),\dots,\zeta(\Delta)]\Big).
\end{align}
\review{The adopted pilot transmission pattern, shown in Figure \ref{pilot table}, is chosen to obtain insightful channel estimators, which highlight the impact of channel time correlation on the system.
This pattern should be interpreted as an analytical abstraction rather than a specific frame design, as practical pilot transmission patterns are often sparse in time and frequency \cite{brunner2024bistatic}.
}
The frequency-domain sensing channel at subcarrier $v$, denoted by $\GG_{i,v} \in 
    \mathbb{C}^{\Mbs \times \Mbs}$, is  defined as  
\begin{align}
    &\GG_{i,v} \triangleq  \sum_{l=1}^L \alpha_l e^{-j2\pi f_{\textrm{D},l}Ti }e^{j2\pi\Delta_f v\tau_l}\mathbf{a}(\theta_l)\mathbf{a}(\theta_l)^\hermit,   
\end{align}
where $\tau_l,\theta_l,f_{\textrm{D},l}$ are the two-way time-delay, the \ac{AoA}, and the Doppler shift of the target, respectively.
\review{Although simple, this point-target model has been found useful for sensing tasks such as "Railway intrusion detection" and "pedestrian detection on highway" \cite{NextGAlliance2025}.}
The Doppler shift of target $l$ is computed as $f_{\textrm{D},l}=2\nu_l/\lambda_c$, with $\nu_l$ 
representing the target's radial velocity.
We assume that the target position changes much more slowly than the communication channel; indeed, its parameters are assumed to be constant for at least $I$ \ac{OFDM} symbols.
The BS is  positioned in the origin, whereas the $l$-th target's position is denoted by $\pp_l=[p_{x,l},p_{y,l}]^\top$; 
it then follows that 
$\theta_l=\atantwo(p_{y,l},p_{x,l}),~ \tau_l=2r_l/c,~ r_l=\norm{\pp_l}.$
Under the narrow-band assumption $\lambda_c \approx \lambda_v$,  
and the steering vector is defined as 
$\mathbf{a}(\theta_l)\triangleq[e^{-j\pi (-\frac{\Mbs}{2}) \sin(\theta_l)} \dots   ~e^{-j\pi (\frac{\Mbs}{2}-1)) \sin(\theta_l)} ]^\top.$
The coefficient $\alpha_l$ is the complex channel gain of target $l$, which is defined as
\vspace{-3mm}
\begin{align}
    \alpha_l=\sqrt{\frac{G_\tx G_\rx c^2 \delta_{\textrm{tg},l}^2 }{(4\pi)^3f_c^2r_l^4}}e^{j2\pi f_c\tau_l},
    \vspace{-3mm}
\end{align}
where $\delta_{\textrm{tg},l}^2$ represents $l$-th target \ac{RCS}. 
Here, $G_\tx,~G_\rx$ are the single-element transmit and receive antenna gains: we assume isotropic single-element antennas, hence $G_\tx,~G_\rx=0$ dBi.
As mentioned in the introduction, in real \ac{ISAC} scenarios, 
the target parameters need to be recovered in an environment dominated by clutter \cite{skolnik2008radar}. 
In the radar literature, clutter is defined as the spurious returns originating from the environment surrounding the targets. 
If not handled properly, clutter causes great harm to the radar system \cite{ward1998space}.
We model clutter as an additive colored complex Gaussian noise term, 
denoted by $\cc_{i,v}$, and notice that if we were to collect the clutter terms on all subcarriers and timeslots in  
$\cc=[\cc_{1,1}^\top,\dots,\cc_{I, V}^\top]^\top$, 
this would be distributed as $   \cc\sim\mathcal{CN}
\left(\mathbf{0},\delta_\textrm{cl}^2\BB\right)$.
\review{
Here, $\delta_\textrm{cl}^2$ represents the clutter's power, which in \ac{SIRV} clutter's model is the texture's second moment.}
\begin{table}[!t]
\begin{center}
\caption {\centering System Parameters}
\begin{tabular}{|m{0.3\linewidth} | m{0.6\linewidth}|} 
 \hline
 \textbf{Notation} & \textbf{Meaning}\\
 \hline\hline 
 $M_\textrm{BS},M_\textrm{UE}$ & Number of BS transmit and receive antennas, UE's receiving antennas\\
 \hline
$S,L$ &  number of transmission streams and sensing targets, respectively\\
 \hline
 $V,\Delta_f, V_\textrm{cho}$ & N$^\circ$ of active subcarriers, subcarrier spacing, and the number of coherent subcarriers\\
 \hline
 $T,T_\textrm{cp},I$  & Symbol time, cyclic prefix duration, and n$^\circ$ of \ac{OFDM} symbols used for radar signal processing\\
 \hline
 $f_c,f_v,\lambda_c,\lambda_v$ & Carrier frequency, $v$-th subcarrier frequency and their wavelengths\\
 \hline
 $\HH_{i},\hh_{i},\CC$ & Channel between the BS and UE at time $i$ in an arbitrary coherence block, $\hh_{i}=\vecc(\HH_{i})$, \\
 \hline
 $\CC,\CC_\bs,\CC_\ue$ & total covariance of $\hh$, $\CC=\CC_\bs\otimes\CC_\ue$ \\
 \hline
 $N_\ue,~\varsigma_\ue$ & Number of clusters of $\hh$ and their angular spread \\
 \hline
$\psi_{\textrm{BS},n},~\psi_{\textrm{UE},n}$ & Median angle of the $n$-th cluster of $\hh$, as seen from the BS and the UE respectively \\
 \hline
 $\zeta(i),~f_{{\rm D}}^\textrm{UE},~\Delta$ & UE's correlation decay coefficient, UE's Doppler shift, and pilot spacing \\
 \hline
 $\GGc_{i,v},\cc_{i,v}$ & Radar channel and observed clutter at subcarrier $v$ and timeslot $i$ \\
\hline
\vspace{0.5mm}
$\theta_l,~\tau_l,~f_{\textrm{D},l},\nu_l,~\alpha_l$ & $l$-th target AoA, round-trip delay, Doppler shift, radial velocity, and channel gain \\
\hline
$\BB_\textrm{sp},~\BB_\textrm{t},~\BB_\textrm{f}$ & Space, time, and frequency clutter covariance matrices \\
\hline
 \vspace{1mm}
$\mathcal{B}_\textrm{sp},~\mathcal{B}_\textrm{t},~\mathcal{B}_\textrm{f}$ & Sample clutter covariance matrices \\
\hline
 \vspace{1mm}
$\widehat{\BB}_\textrm{sp},~\widehat{\BB}_\textrm{t},~\widehat{\BB}_\textrm{f}$ & Estimates of the clutter covariance matrices \\
\hline
\vspace{1mm}
$\overline{\BB}_\textrm{f},\widetilde{\BB}_\textrm{f}$ & Discrete and diffusive part of the clutter's frequency covariance \\
\hline
$N,~\psi_n,~\Tilde{r}_n~,f_{\textrm{D},n}$ & Number of clutter patches, $n$-th patch median angle, range, and Doppler shift, respectively\\
\hline
$\varsigma_\textrm{sp},~\varsigma_\textrm{D},~\varsigma_\tau  $  & Clutter patches angular, Doppler, and delay spreads, respectively\\
\hline
$\mu_f,~B_\textrm{cho},~\chi  $  &\vspace{0.5mm} Characterization of $\widetilde{\BB}_\textrm{f}$: exponential base, coherence bandwidth, and power coefficient \\
\hline
 $\ff_{i,v,s},x_{i,v,s},\rho_{i,v,s}$ & Precoding vector, power allocation coefficient, and transmitted symbol on the $s$-th stream, $v$-th subcarrier, $i$-th timeslot \\
 \hline
$\nnb_{i,v},\nn_{i,v},\sigma^2_\ue,\sigma^2$ & UE and radar receiver noises and respective powers\\
 \hline
 $\alpha_{l},\alpha_\textrm{UE}$ & $l$'th target complex channel gain and UE's pathloss\\
 \hline
 $\tau_\rmp,\boldsymbol{\beth}$ & Number of subcarriers used for pilot transmission and pilot matrix\\
 \hline
 \vspace{1mm}
 $\widehat{\hh}_{i},\,\widetilde{\hh}_{i},\,\widehat{\boldsymbol{\Xi}}_{i},\,\widetilde{\boldsymbol{\Xi}}_{i} $ & estimate and estimation error of $\hh_{i}$ and their respective covariance matrices.\\
\hline
$\boldsymbol{\phi}_{i,v,s}, \hat{\boldsymbol{\phi}}_{i,v,s}$ & precoded radar channel at stream $s$ and its estimate.\\
 \hline
 \end{tabular}
 \end{center}
\end{table}
\review{
Within the \ac{SIRV} framework, the texture is usually a random variable with a heavy-tailed distribution (e.g., Weibull) 
as it represents slow variations of the clutter energy \cite{duan2024analytical}. 
For the sake of mathematical tractability, we set this parameter to a constant value:   this choice preserves the Gaussianity 
of $\cc$, 
making the subsequent mathematical derivations more insightful.
It should be noted that a stochastic texture would not invalidate the covariance-based clutter processing presented in the following sections, as the texture influences the clutter covariance only through its second moment\cite{greenewald2016robust}. 
}
The transmitted waveform during a generic data-timeslot $i$ at subcarrier $v$ is defined as 
\begin{align}
\ssb_{i,v}=
\sum_{s=1}^S\underbrace{\ff_{i,v,s}\sqrt{\rho_{i,v,s}}x_{i,v,s}}_{\ssb_{i,v,s}}=\FF_{i,v}\PP_{i,v}\xx_{i,v} \in \mathbb{C}^{\Mbs},
\end{align}
where $\PP_{i,v}=\diag(\boldsymbol{\rho}_{i,v})$ and $ \FF_{i,v} \in \mathds{C}^{\Mbs \times S}$ represents the precoding matrix. 
\review{The adopted precoding structure follows a functional separation approach, where communication and sensing are served by separate precoding streams, each designed to maximize the performance of its relative task.
We allocate $\Mue$ streams to the UE and an additional one for sensing purposes, bringing the total number of streams to $S=\Mue+1$. } The precoding matrix associated to the UE is
$\FF_{i,v}^\ue=[\ff_{i,v,1}^\ue,\dots,\ff_{i,v,\Mue}^\ue]\in\mathbb{C}^{\Mbs \times \Mue}$, thus making the total precoding matrix equal to
$\FF_{i,v} \triangleq [\FF_{i,v}^\ue,\ff_{i,v,S}]\in\mathbb{C}^{\Mbs \times S}$.
We assume that each column of that matrix is normalized to have a unitary norm. 
We denote by $\rho_{i,v,s}$ the power assigned by the BS 
to the $s$-th stream on subcarrier $v$ at time $i$. 
More specifically $\boldsymbol{\rho}_{i,v}\triangleq \hspace{-1mm}\left[\boldsymbol{\rho}_{i,v}^{\ue,\top}~\sqrt{\rho_{i,v,S}}\right]^\top$, with $\boldsymbol{\rho}_{i,v}^\ue=\left[\sqrt{\rho_{i,v,1}^\ue}\dots\sqrt{\rho_{i,v,\Mue}^\ue}\right]^\top$.
The total power budget $P_\tx$ of the \ac{BS} is divided between sensing and communication, where the individual power budgets are denoted by $P_{\textrm{comm}}$ and $P_{\textrm{sens}}$. 
The BS employs the entirety of its power budget during each timeslot $i$, thus $P_{\textrm{comm}}$ and $P_{\textrm{sens}}$ can be defined as  
\begin{align}
    &P_{\textrm{comm}} = \sum_{s=1}^{\Mue}\sum_{v=1}^v \rho_{i,v,s}=P_\tx\gamma,~\forall i,\\
    &P_{\textrm{sens}} = \sum_{v=1}^v \rho_{i,v,S}=P_\tx(1-\gamma),~\forall i.
\end{align}
\review{Here, $\gamma \in [0,1]$ represents the percentage of the total available power dedicated to communication. This parameter embodies the inherent power trade-off between sensing and communications.  } 
At time instance $i$, 
the transmitted symbols on subcarrier $v$, either pilots or information-bearing symbols, 
are defined as $\xx_{i,v}=[\xx_{i,v}^{\ue,\top} ,x_{v,i,S}]^\top$,
 where $\xx_{i,v}^\ue \in \mathds{C}^{\Mue}$ is the vector of symbols meant for the \ac{UE} and $x_{i,v,s} \sim 
\mathcal{CN}(0,1)$. 
The frequency-domain observation at the UE during a generic data slot $i$ at subcarrier $v$ is defined as 
\begin{align}\label{y generic data}
\yyb_{i,v}\triangleq\alpha_\ue\HH_{i,k}\ssb_{i,v} + \nnb_{i,v} ~\in \mathbb{C}^{\Mue},
\end{align}
where $\nnb_{i,v}$ is the receiver noise at the UE at timeslot $i$ and subcarrier $v$, 
whose entries are i.i.d. random variables distributed as  
$\mathcal{CN}(0,\sigma_{\ue}^2)$.
On the other hand, during any timeslot $i$, the observed radar echo at the \ac{BS} can be defined as 
\begin{align}\label{y radar}
\yy_{i,v}\triangleq\boldsymbol{\mu}_{i,v}+\cc_{i,v}+\nn_{v,i}^\rmr~\in \mathbb{C}^{\Mbs}.
\end{align}
where $\boldsymbol{\mu}_{i,v}=\GGc_{i,v}\ssb_{i,v}$ represents the noiseless and clutter-less observation.
Once again, $\nn_{i,v}^\rmr$ is the radar receiver noise at timeslot $i$ and subcarrier $v$, whose entries are i.i.d. random variables distributed as $\mathcal{CN}(0,\sigma^2)$.
\vspace{-2mm}

\section{Characterizing the Impact of Channel Aging and Clutter}\label{cov mod sec}
\begin{figure}[t!]
\begin{center}
   \resizebox{0.5\textwidth}{!}{
    \includegraphics[]{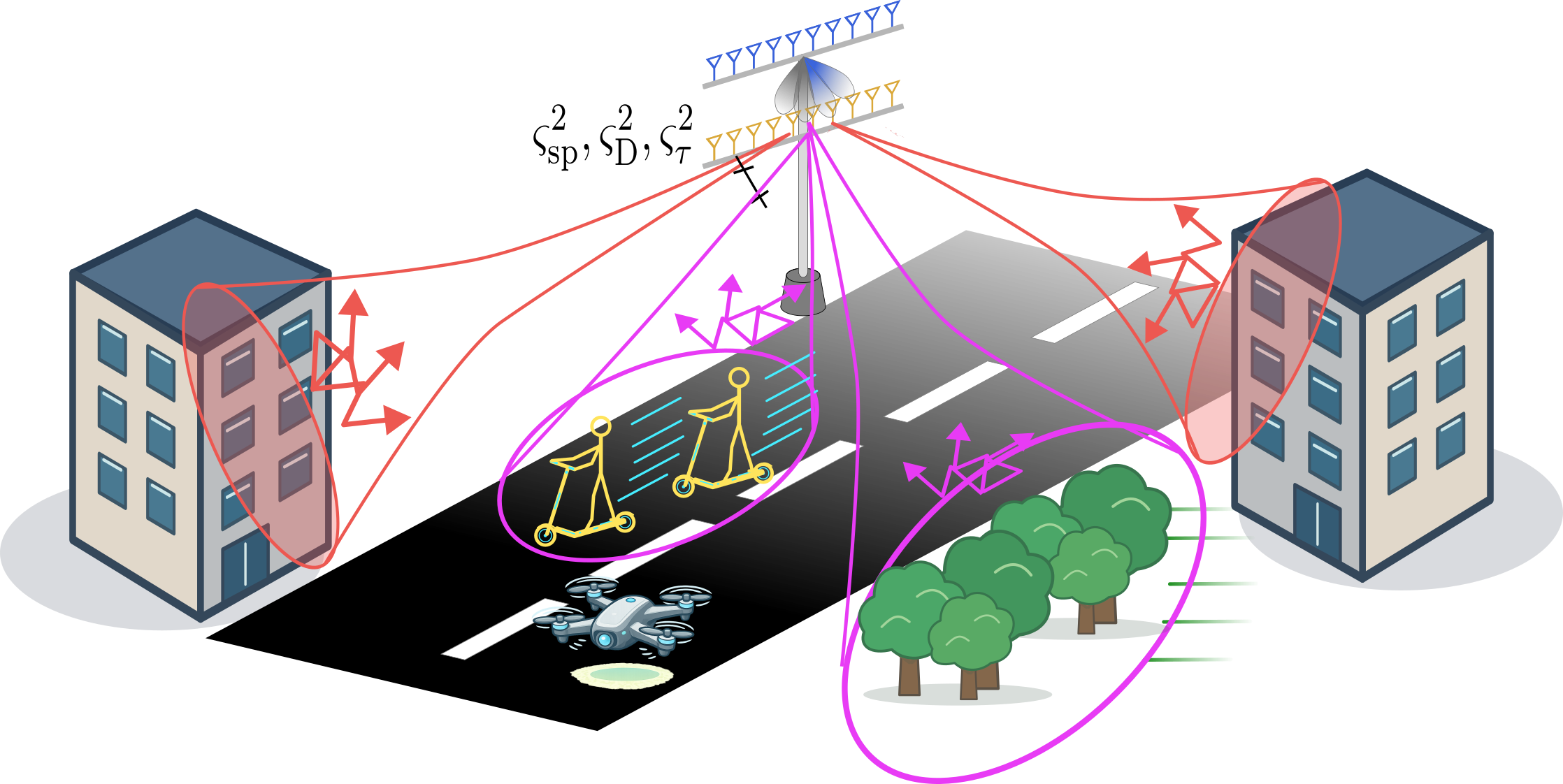}}
      \vspace{-3mm}
	 \caption{ \review{Conceptual representation of an urban clutter environment where Kronecker separability provides a tractable approximation. Clutter is formed by the superimposition of a discrete number of clutter patches, both static and dynamic. 
     }}
    \label{clutter geom}
		\end{center}
  \vspace{-4mm}
\end{figure}

In this section, we characterize the covariances of the communication channels and the observed clutter. 
\review{Inspired by  the 3GPP standardized cluster delay-line channel \cite{38901}, the spatial covariance matrix of $\hh_{i}$ 
follows a non-isotropic multipath model, originated by a superimposition of clusters, each scattering multiple rays with a narrow angle spread.}
Fig.~\ref{clutter geom} shows how clutter is generated by approximately independent clutter patches, each contributing to the total covariance.
Recall that a clutter patch refers to 
a region of the environment that produces radar echoes that approximately have the same angle, 
Doppler shift, and range; indeed, we assume that the spreads describing the patch 
(i.e., angle, Doppler, and delay spread) are small.
\review{
It is important to keep in mind that, although the same mathematical model describes clutter patches and communication clusters, they represent different physical scales.
Specifically, a clutter patch represents an extended surface while a UE cluster is a localized multipath component, which captures the effect of small-scale fading close to the UE. For this reason, the set of clusters creating $\CC$ 
does not constitute a clutter patch. However, the model presented in this section can be easily generalized to scenarios where the same object contributes to clutter and UE multipath.}


\subsection{Spatial and Temporal Covariances of UE Channels}
We will now focus on the computation of the BS-side correlation $\CC_{\bs}$, as the procedure to retrieve $\CC_{\ue}$ is entirely analogous.
In its most general form, we may write that 
\begin{align}
    \CC_{\bs} = \int_{-\frac{\pi}{2}}^{\frac{\pi}{2}} f_\bs(\psi)\atx(\psi)\atx(\psi)^\hermit d\psi,
\end{align}
where $f_\bs(\psi)$ is the angular spreading function, which specifies the gain and phase shifts associated with every angular direction.
The \ac{LoS} between BS and the UE is blocked, and the signal transmitted by the BS reaches the UE through a set of $N_\ue$ clusters\cite{demir2022RISP}.
The angular spreading function can be modeled as \cite{abdi2002space} 
\begin{align}
    f_\bs(\psi) = \frac{1}{N_\ue}\sum_{n=1}^{N_\ue} \frac{1}{\sqrt{2\pi}\varsigma_\ue} \exp\left\{-\frac{(\psi- \psi_{\textrm{BS},n} )^2}{2\varsigma_\ue^2}\right\},
\end{align}
where $\psi_{\textrm{BS},n}$ is the $n$-th cluster angle as seen from the BS: all the clusters are seen by the BS within an angular interval $\Delta\psi_{\textrm{BS},k}$.
Each cluster reflects the impinging signal with an angular spread of $\varsigma_\ue$. By assuming small angular spreads  i.e., $\cos(\varsigma_\ue)\approx1$ and $\sin(\varsigma_\ue)\approx \varsigma_\ue$, each entry of  $\CC_{\bs}$ can be tightly approximated as \cite[Lemma 2]{demir2022channel}
    \begin{align}\label{C_k_eq}
&\hspace{-2mm}\left[\CC_{\bs}\right]_{l,m}\hspace{-1mm}\approx\hspace{-2mm}\sum_{n=1}^{N_\ue}\frac{A_{l,m}(\psi_{\normalfont{\textrm{BS}},n})}{N_\ue} e^{-\frac{\pi^2}{2}( (l-m)\cos(\psi_{\normalfont{\textrm{BS}},n})\varsigma_\ue)^2},
    \end{align}
    where $~l,m=1,\dots, \Mbs,$ and $A_{l,m}(\psi_{\normalfont{\textrm{BS}},n})= e^{-j\pi (l-m)\sin(\psi_{\normalfont{\textrm{BS}},n})}$.
The temporal evolution (aging)
of the channel is modeled as a first-order \ac{AR} process with a scale-identity state-transition matrix. 
The scaling coefficient $\zeta(i)$ representing the correlation decay is chosen as $ \zeta(i)=J_0(2\pi T f_{{\rm D}}^\textrm{UE} i),$
where $f_{{\rm D}}^\textrm{UE}$ is the UE's Doppler shift and $J_0(\cdot)$ is the zero-th order Bessel function\cite{wang2003efficient}.
The dependence of $\zeta $ on $f_{{\rm D} }^\textrm{UE}$  is motivated by the latter's direct impact on the coherence time of $\hh $, thus constituting a good indication of the time correlation decay.
\review{In the remainder of the paper, we shall respectively denote with $\psi_{\ue,n}$ and $\Delta\psi_{\ue}$ the angle of the $n$-th cluster and the angular span occupied by the clusters as seen by the UE.}

\subsection{Clutter Characterization}
Recall that clutter originates from a few strong patches (Fig. \ref{clutter geom}), creating reflected echoes with ``narrow" lobes. 
Under this assumption, the clutter covariance matrix can be approximated as  
\begin{align}
    &\mathds{E}[\cc \cc^\hermit]= 
    \delta_\textrm{cl}^2\BB \approx \sqrt[3]{\delta_\textrm{cl}^2}\BB_\textrm{sp} \otimes \sqrt[3]{\delta_\textrm{cl}^2}\BB_\textrm{t} \otimes \sqrt[3]{\delta_\textrm{cl}^2}\BB_\textrm{f}. 
\end{align}
This model is widely studied in the radar literature and is referred to as the Kronecker-separable clutter model\cite{greenewald2016robust,greenewald2013kronecker, Werner:08}.
\review{This structure should be intended as an approximation, which is adopted as an enabler for computationally efficient signal processing operations.
A scenario for which Kronecker-separability is a good approximation is the so-called street-canyon, i.e., a large boulevard with rows of buildings on each side, as clutter is dominated by extended planar surfaces, such as building facades.
Here, clutter echoes produce angularly localized echoes, with weak Doppler-delay coupling.
On the other hand, this patched behavior can be seen in \cite{henninger2023crap} and \cite{henninger2024crap2}: two different experimental setups have shown how clutter presents a dominant component which can be modelled as a clutter patch.
}
Here, $\BB_\textrm{sp}$, $\BB_\textrm{t}$, and $\BB_\textrm{f}$ model the clutter's angle, Doppler, 
and range distribution.
The main advantage of this model is its low computational complexity, stemming from the fact that operations such as whitening, eigen-decomposition, and inversion can be applied independently to each dimension.
The space covariance $\BB_\textrm{sp}$ is computed in the same way as in \eqref{C_k_eq}, as clutter patches and UE clusters have the same mathematical model.
Let us denote the patches' angular spread as $\varsigma_{\normalfont{\textrm{sp}}}$. Similarly to \eqref{C_k_eq}, by assuming $\cos(\varsigma_{\normalfont{\textrm{sp}}})\approx1$ and $\sin(\varsigma_{\normalfont{\textrm{sp}}})\approx \varsigma_{\normalfont{\textrm{sp}}}$ ,
each entry f $\BB_{\normalfont{\textrm{sp}}}$ is approximated by
    \begin{align}\label{B_s_eq}
&\left[\BB_{\normalfont{\textrm{sp}}}\right]_{l,m}\hspace{-1mm}\approx \frac{1}{N}\sum_{n=1}^N A_{l,m}(\psi_n) e^{-\frac{1}{2}(\pi (l-m)\cos(\psi_n)\varsigma_\textrm{sp} )^2}, 
    \end{align}
where $~l,m=1,\dots, \Mbs,$ and $A_{l,m}(\psi_n)= e^{-j\pi (l-m)\sin(\psi_n)}$.
The role of the temporal covariance matrix differs from that in communication-oriented analyses. In radar applications, the matrix
$\BB_{\normalfont{\textrm{t}}}$ captures the clutter's Doppler signature, allowing for distinguishing it from the target's. For this reason, $\BB_{\normalfont{\textrm{t}}}$ 
is computed similarly to $\BB_{\normalfont{\textrm{sp}}}$.
Let us start by defining each element $\BB_{\normalfont{\textrm{t}}}$ as the inverse Fourier transform of the power spectral density
\begin{align}\label{Bt int}
\BB_{\normalfont{\textrm{t}}} = \int_{-\infty}^{+\infty}g(f)\bb(f)\bb(f)^\hermit  df,
\end{align}
where $\bb(f)$ is the Doppler steering vector
   $  \bb(f)=[1~e^{-j2\pi f_{\textrm{D},n} T }~\dots~e^{-j2\pi fT (I-1) }]^\top \in \mathds{C}^{I}.$
Here, $g(f)$ is the power spectral density function, specifying the gain associated with each frequency component. 
Once again, this function is modeled as the summation of independent patches, that is 
\begin{align}\label{gf}
    g(f)=\frac{1}{N}\sum_{n=1}^{N} \frac{1}{\sqrt{2\pi}\varsigma_\rmd} \exp\left\{-\frac{(f- f_{\rmd,n} )^2}{2\varsigma_\rmd^2}\right \}.
\end{align}
Here, $f_{\rmd,n}$ is the $n$-th patch Doppler shift and $\varsigma_\rmd^2$ is the Doppler spread.
The clutter's Doppler spread $\varsigma_\textrm{D}$ can be further expanded into $\varsigma_\textrm{D}=1/2\pi T I_\textrm{c}$, where $I_\textrm{c}$ represents the number of \ac{OFDM} symbols for which the clutter time correlation is greater than $0$. 
\begin{customlemma}{1}\label{lemma bt}
 Under the assumption of almost independent patches, each entry of $\BB_{\normalfont{\textrm{t}}}$ is computed as 
    \begin{align}\label{B_t_eq}
&\left[\BB_{\normalfont{\textrm{t}}}\right]_{l,m}\hspace{-1mm}=\frac{1}{N} \sum_{n=1}^N D_{l,m}(f_{\rmd,n}) e^{-\frac{(l-m)^2}{2I_c^2}}, 
    \end{align}
    where $~l,m=1,\dots, I,$ and $D_{l,m}( f_{\textrm{D},n})= e^{-j2\pi f_{\textrm{D},n} T (l-m)}$.
\end{customlemma}
\begin{IEEEproof}
Let us substitute \eqref{gf} into \eqref{Bt int} to obtain  $\left[\BB_{\normalfont{\textrm{t}}}\right]_{l,m} = \frac{1}{N}\sum_{n=1}^N\int_{-\infty}^{+\infty}\frac{1}{\sqrt{2\pi}\varsigma_\rmd} \exp\left\{-\frac{(f- f_{\rmd,n} )^2}{2\varsigma_\rmd^2}\right \}e^{-j2\pi fT (l-m) }  df$. By applying the square completion method and taking out of the integrals the terms constant in $f$, we can then rewrite the previous expression as
\begin{align}
&\left[\BB_{\normalfont{\textrm{t}}}\right]_{l,m} =\frac{1}{N}\sum_{n=1}^Ne^{-j2\pi f_{\textrm{D},n} T (l-m)} e^{-2(\pi T (l-m)\varsigma_\textrm{D})^2}\times \nonumber\\
&\int_{-\infty}^{+\infty}\frac{1}{\sqrt{2\pi}\varsigma_\rmd} e^{-\frac{(f-f_{\textrm{D},n}-j2\pi \varsigma_{\normalfont{\textrm{D}}}^2T(l-l) )^2}{\varsigma_\textrm{D}^2} } df.    
\end{align}
The last integral integrates to 1 as $\varsigma_{\normalfont{\textrm{D}}} >0$, by substituting $\varsigma_\textrm{D}=1/2\pi T I_\textrm{c}$ the expression follows.
\end{IEEEproof} 

The frequency covariance matrix is the Fourier transform of the power-delay spectrum
\begin{align}
    \BB_{\normalfont{\textrm{f}}} = \int_{-\infty}^{+\infty}h(\tau)\dd(\tau)\dd(\tau)^\hermit  d\tau,
\end{align}
where $\dd(\tau)$ is the delay steering vector, defined as 
    $\dd(\tau)=[1~e^{j2\pi \Delta_f \tau }~\dots~e^{2\pi \Delta_f \tau (V-1) }]^\top \in \mathds{C}^{V}.$
Here, $h(\tau)$ is the power delay profile, associating a gain to a delay.
Unlike a target, each clutter patch is an extended surface composed of multiple scatterers; this spoils the range coherence of each patch, as the surface creates a diffusive reflection. The back-scattered echo from each of these patches is also affected by multipath. For this reason, in addition to a "discrete" term similar to the previously shown ones, $h(\tau)$ contains an additional term, denoted by $\tilde{h}(\tau) $, representing multipath and diffusive reflections.
The power delay profile is thus defined as 
\begin{align}
     h(\tau)=\frac{1}{N}\sum_{n=1}^{N} \frac{1}{\sqrt{2\pi}\varsigma_\tau} \exp\left\{-\frac{(\tau- \Tilde{\tau}_n )^2}{2\varsigma_\tau^2}\right \} +  \tilde{h}(\tau) ,
\end{align}
where $\tilde{r}_n$ is the median range of the $n$-th patch and $\tilde{\tau}_n=2\tilde{r}_n/c$.
Because of this modeling choice  $\BB_{\normalfont{\textrm{f}}}= \overline{\BB}_{\normalfont{\textrm{f}}} + \widetilde{\BB}_{\normalfont{\textrm{f}}}$.
Let us denote with $\varsigma_\tau$ the patch's delay spread, to obtain:

\begin{customlemma}{2}\label{lemma bf}
 Under the assumption of almost independent patches, each entry of $\BB_{\normalfont{\textrm{f}}}$ is computed as 
    \begin{align}\label{B_f_eq}
&\left[\overline{\BB}_{\normalfont{\textrm{f}}}\right]_{l,m}\hspace{-1mm}=\frac{1}{N} \sum_{n=1}^N E_{l,m}(\Tilde{r}_{n}) e^{-2(\pi \Delta_f (l-m)\varsigma_\tau )^2}; \\
&\widetilde{\BB}_{\normalfont{\textrm{f}}}=\chi\mathcal{T}\left(1 ~\mu_{\normalfont{\textrm{f}}}^{\frac{\Delta_f}{B_\textrm{cho}}}     \dots
\mu_{\normalfont{\textrm{f}}}^{\frac{(V-1)\Delta_f}{B_\textrm{cho}}}
\right),
    \end{align}
    where $~l,m=1,\dots, V,$ and $E_{l,m}( \Tilde{r}_{n})= e^{j2\pi \Delta_f \frac{2\Tilde{r}_{n}}{c}  (l-m)}$.
\end{customlemma}
\begin{IEEEproof}
$\overline{\BB}_{\normalfont{\textrm{f}}}$ is obtained as in Lemma \ref{lemma bt}. On the other hand, we assume that the progression along $\tau$ of $\tilde{h}(\tau)$ follows an AR1 structure, making its contribution to $\BB_{\normalfont{\textrm{f}}}$  a Toeplitz matrix \cite{sherman2023eigenstructure}.
\end{IEEEproof}

Here, $\widetilde{\BB}_{\normalfont{\textrm{f}}}$ is characterized by  $\mu_{\normalfont{\textrm{f}}} < 1$, which is the correlation's decay magnitude, the coherence bandwidth $B_\textrm{cho}$, and $\chi$, which is the power of the diffusive clutter w.r.t. the coherent one.
\review{
\subsection{Robustness to Non-separable Clutter}\label{rob sec}
The previously introduced Kronecker-separable assumption provides a tractable mathematical model for clutter and enables efficient mode-wise operation. However, in real environments, coupling among clutter dimensions may arise from heterogeneous clutter patches, calibration errors, or non-stationarity in the clutter itself.
For this reason, we assess the robustness of the proposed radar processing to non-separable clutter by introducing a controllable deviation from the separable model assumed so far.
More specifically, the clutter covariance is now a linear combination of the previously presented separable component and a perturbing one. 
The stacked clutter vector $\cc$ is then distributed as 
\begin{align}
\cc &\sim
\mathcal{CN}\big(\mathbf{0},(1-\epsilon)\delta_\textrm{cl}^2\BB + \epsilon \ZZ\big),
\end{align}
where $\epsilon$ is the clutter's degree of non-separability.
The matrix $\ZZ$ is once again defined as a Kronecker product, that is $\ZZ=\ZZ_\textrm{sp} \otimes \ZZ_\textrm{t} \otimes  \ZZ_\textrm{f}$. This choice allows us to maintain the computationally efficient generation of clutter samples while simulating a non-separable clutter covariance, thanks to the property $(\AAb \otimes \BB) + (\CC \otimes \DD) \neq (\AAb + \CC) \otimes (\BB +\DD)$.
Each of the matrices composing $\ZZ$ is obtained by applying an eigenvector rotation and eigenvalue perturbation onto its nominal counterpart. This perturbation model allows us to individually investigate the impact of mismatches in both the power distribution and the underlying subspace structure.
Therefore, $\ZZ_\textrm{t}$, and analogously $\ZZ_\textrm{sp}$ and $\ZZ_\textrm{f}$, is defined as 
\begin{align}
    \ZZ_\textrm{t}= \boldsymbol{\Pi}_\textrm{t}\UU_\textrm{t}\left(\boldsymbol{\Sigma}_\textrm{t} + \eta_\textrm{val}\widetilde{\boldsymbol{\Sigma}}_\textrm{t} \right)\UU_\textrm{t}^\hermit\boldsymbol{\Pi}_\textrm{t}^\hermit \in \mathds{C}^{I \times I},
\end{align}
where $\UU_\textrm{t}, \boldsymbol{\Sigma}_\textrm{t}$ denote the eigen decomposition of $\sqrt[3]{\delta_\textrm{cl}^2}\BB_{\normalfont{\textrm{t}}}(\sqrt[3]{\delta_\textrm{cl}^2}\BB_{\normalfont{\textrm{f}}})$
. Here $\eta_\textrm{val}$ denotes the strength of the eigenvalue perturbation and $\widetilde{\boldsymbol{\Sigma}}_\textrm{t}$ is a diagonal matrix whose entries are complex Gaussian distributed with unitary variance. 
The eigenvalue rotation matrix is defined as $\boldsymbol{\Pi}_\textrm{t}=\exp\{ \eta_\textrm{vec} \widetilde{\boldsymbol{\Pi}} \}$
, with $\widetilde{\boldsymbol{\Pi}}$ being a skew-hermitian matrix and $\eta_\textrm{vec}$ denoting the strength of this eigenvector rotation.
}


\section{Communication Channel Estimation}\label{ch est sec}
We now provide a channel estimator for the communication channel.
\review{ 
As stated in Sec. \ref{sys mod sec}, we operate under the instantaneous UE-feedback assumption, granting estimated \ac{CSIT} in addition to \ac{CSIR}. 
This allows us to focus on studying the gain of known temporal correlation over block fading, without explicitly modelling any feedback mechanism.
In practice, \ac{CSIT} can be achieved by scheduling an additional uplink pilot slot after the downlink one \cite {kama2024downlink}: under moderate \ac{UE} mobility, the \ac{UL} and \ac{DL} estimates will follow a very similar trajectory.
We reserve the implementation of this feedback mechanism for future work.
}
It is also important to mention that the UE's time and space covariance matrices are assumed to be known. The UE is associated with an $S$-stream pilot matrix denoted by $\boldsymbol{\beth}\in\mathbb{C}^{S \times \tau_\rmp}$:  $\tau_\textrm{p}$ subcarriers of each coherence block are used to estimate the UE's channel. 
We assume that during pilot transmission, the BS uses the same precoding matrix for all the subcarriers within the same block; we can thus drop the $v$ index for the rest of this sub-section and derive the channel estimator of an arbitrary coherence block.
Let us consider the arbitrary pilot slot $\iota$, the signals received by the UE on all the $\tau_\rmp$ subcarriers used for channel estimation can be written as 
\begin{align}
    \YYb_{\iota} = \alpha_\ue\HH_{\iota}\FF_{\iota}\PP_{\iota}\boldsymbol{\beth} 
    + \NNb_\iota\in \mathbb{C}^{\Mue \times \tau_\rmp},
\end{align}
where $\NNb_\iota=[\nnb_{\iota,1},\dots,\nnb_{\iota,\tau_\rmp}]$.
To ensure a correct elimination of the pilot from the received signal, $\boldsymbol{\beth}$ must satisfy $\boldsymbol{\beth}\boldsymbol{\beth}^\hermit = \tau_\rmp\mathbf{I}_S$: this can be achieved when $\tau_\rmp \geq S$.
Precoded transmission during channel estimation is adopted to reduce the minimum $\tau_\rmp$, since without precoding $\tau_\rmp \geq \Mbs$ \cite{he2024mse}.
The UE de-spreads this observation as $\widetilde{\YYb}_{\iota}\hspace{-1mm}\triangleq\YYb_{\iota}\boldsymbol{\beth}^\hermit/\sqrt{\tau_\rmp}=\alpha_\ue\sqrt{\tau_\rmp}\HH_{\iota}\FF_{\iota}\PP_{\iota}+ \widetilde{\NNb}_{\iota},$
where $\widetilde{\NNb}_{\iota}=\NNb_{\iota}\boldsymbol{\beth}^\hermit\in \mathbb{C}^{\Mue \times S}$. Let us introduce 
\begin{align}
\widetilde{\yyb}_{\iota}=\vecc(\widetilde{\YYb}_{\iota})=
\alpha_\ue\sqrt{\tau_\rmp}\breve{\FF}_{\iota}\hh_{\iota} +  \widetilde{\nn}_{\iota} ~\in \mathbb{C}^{S\Mue},
\end{align}
where $\breve{\FF}_{\iota}=(\FF_{\iota}\PP_{\iota})^\top \otimes \mathbf{I}_{\Mue} \in \mathds{C}^{S\Mue \times \Mue \Mbs }$ and $\widetilde{\nnb}_{\iota}=\vecc(\widetilde{\NNb}_{\iota})$.
The estimate of the channel at the receiver can be based on multiple pilot observations before the actual data slot $i$.
We assume that the receiver uses a total of $p$ pilot observations before the present one, bringing the total number of observations to $p_\textrm{tot}=p+1$, which are encompassed in $  \widetilde{\yyb}_{\ptot}\hspace{-1mm}=
    [\widetilde{\yyb}_{\iota }^\top,\dots\widetilde{\yyb}_{\iota-(p_\textrm{tot}-1)(\Delta+1)}^\top]^\top\hspace{-1mm}\in \mathbb{C}^{S\Mue p_\textrm{tot}}.$
The following lemma shows how a channel estimate of $\HH_{i}$, with $\iota$ being the closest pilot, is obtained.
\begin{customlemma}{3}\label{lemma hk}
    The \ac{MMSE} channel estimate of $\hh_{i}$ given the pilot spacing $\Delta$ is obtained as
    \begin{align}
    &\widehat{\hh}_{i} = \AAb_{i}\hspace{0.5mm}\widetilde{\yyb}_{\ptot} ~\in \mathbb{C}^{\Mbs\Mue},
     \label{hmmse}
     \end{align}
where
\begin{align}
&\AAb_{i}= \alpha_\ue \sqrt{\tau_\rmp}\hspace{1mm}\EE_{i}\breve{\FF}_{p_\textrm{tot}}^\hermit\hspace{0.5mm}\overline{\AAb}_{i}^{-1},\\
&\overline{\AAb}_{i}=\alpha_\ue^2\tau_\rmp \breve{\FF}_{p_\textrm{tot}}\MM_i\breve{\FF}_{p_\textrm{tot}}^\hermit + \sigma^2_\ue\mathbf{I}_{\ptot S\Mue},\\
&\breve{\FF}_{p_\textrm{tot}} = \normalfont{\blkdiag}\left(\breve{\FF}_{\iota},\dots,\breve{\FF}_{\iota-(\ptot -1)(\Delta+1)} \right),\\
&\EE_{i}=\left[\zeta(\overline{i}),\dots,\zeta(\overline{i}+(\ptot -1)(\Delta+1))  \right] \otimes \CC,\\
&\MM_{i}\hspace{-1mm}= \mathcal{T}\left([\zeta(0),\dots,\zeta((\ptot -1)(\Delta+1))]\right)\otimes \CC ,
\end{align}
and $\overline{i}=i-\iota$.
\end{customlemma}
\begin{IEEEproof} 
Let us start by rewriting  $ \widetilde{\yyb}_{\ptot}$ as 
\begin{align}
\widetilde{\yyb}_{\ptot}=\alpha_\ue\sqrt{\tau}\breve{\FF}_{p_\textrm{tot}}\hh_{ \ptot} +
\nnb_{\ptot} \in \mathds{C}^{S\Mue\ptot},
\end{align}
where
\begin{align}
&\hh_{\ptot}=[\hh_{\iota, k}^\top,\dots\hh_{\iota-(p_\textrm{tot}-1)(\Delta+1)}^\top]^\top \in \mathds{C}^{\Mue \Mbs \ptot};\\
&\nnb_{\ptot}=[\nnb_{\iota}^\top,\dots\nnb_{\iota-(p_\textrm{tot}-1)(\Delta+1)}^\top]^\top \in \mathds{C}^{S \Mue \ptot}.
\end{align} 
The canonical form of the MMSE estimator of $\hh_{i}$ is \cite{kay2013fundamentals} 
\begin{align}
    \widehat{\hh}_{i}=\mathds{E}[\hh_{i}\widetilde{\yyb}_{\ptot}^\hermit]\mathds{E}[\widetilde{\yyb}_{\ptot}\widetilde{\yyb}_{ \ptot}^\hermit]^{-1}\widetilde{\yyb}_{\ptot}.
\end{align}
Then, the expectations are easily computed by leveraging the known temporal correlation structure\cite{fodor2021performance}, indeed $\mathds{E}[\hh_{i}\widetilde{\yyb}_{ \ptot}^\hermit]= \alpha_\ue \sqrt{\tau_\rmp}\hspace{1mm}\EE_{i}\breve{\FF}_{p_\textrm{tot}}^\hermit$ and $\mathds{E}[\widetilde{\yyb}_{ \ptot}\widetilde{\yyb}_{k,\ptot}^\hermit]=  \overline{\AAb}_{i}$. By substituting the expectations in $\widehat{\hh}_{i}$ the lemma follows. 
\end{IEEEproof}
We move on to characterizing the distribution of the channel estimate
\begin{customcor}{1}\label{corollary xi}
The channel estimate $\widehat{\hh}_{i}$ is a zero-mean complex Gaussian random vector $\widehat{\hh}_{i}\sim \mathcal{CN}(\mathbf{0},\widehat{\boldsymbol{\Xi}}_{i})$, where $\widehat{\boldsymbol{\Xi}}_{i,k}= \alpha_\ue^2\tau_\rmp \EE_{i,k}\breve{\FF}_{p_\textrm{tot}}^\hermit\overline{\AAb}_{i}^{-1}\breve{\FF}_{p_\textrm{tot}}\EE_{i}^{\hermit}.$
\end{customcor}
Stemming from  Lemma \ref{lemma hk} and Corollary \ref{corollary xi}, we can now characterize the channel estimation error.
\begin{customcor}{2}
 The channel estimation error $\widetilde{\hh}_{i}=\widehat{\hh}_{i}-\hh_{i}$ is a complex Gaussian zero-mean vector whose covariance matrix can be defined as $    \widetilde{\boldsymbol{\Xi}}_{i}=\CC-\widehat{\XXi}_{i} \in \mathbb{C}^{\Mbs\Mue \times \Mbs\Mue}.$  
\end{customcor}
\review{It should be noted that the present estimator could be easily extended to accommodate an \ac{AR} aging structure over subcarriers. 
Under the assumption of separable time-frequency aging, the extension follows from the covariance-based formulation presented in Lemma \ref{lemma hk}: this extended estimator would then be able to operate with pilot patterns that are sparse in both time and frequency.
}

\section{Radar signal processing} \label{sec pipeline}
\reviewAfter{This section presents the clutter estimation and suppression procedure. Starting from the sensing observation,
the BS estimates the clutter second-order statistics and uses them to suppress its contribution. Then RA and RV maps are computed: these maps serve as a sensing performance representation, upon which detection and estimation algorithms may be applied.}
\footnote{For analytical tractability, $N,\delta^2_\textrm{cl}, \varsigma_\textrm{sp},\varsigma_\textrm{D},$ and $\varsigma_\tau$ are assumed to be known. These quantities can be reliably estimated from target-free range–Doppler bins; thus, this assumption does not limit the practical applicability of the proposed pipeline.} 
The first step is the estimation of the clutter's second-order statistics, so that the action of the former can be eliminated: this is carried out by leveraging the assumed sparsity in each dimension.
Secondly, the action of the transmitted symbols is eliminated through clutter-aware \ac{MF}. 
The cleaned data is then used to compute \ac{RA} and \ac{RV} maps associated with every stream.
Thanks to the slowly-changing target assumption, $I$ timeslots are used for radar signal processing tasks.
\reviewAfter{Once the maps have been obtained, estimation and detection may be applied to them \cite{liyanaarachchi2023joint}; 
We have here presented range maps only to assess the clutter suppression capabilities and the target identifiability. The target peaks' sharpness, location, and magnitude should be interpreted as an indication of the estimation and detection performance of the system rather than their substitutes.
An explicit mathematical derivation of a parameter estimator and a target detector is left for future work.
}

\subsection{Clutter Covariance Estimator}\label{cl est sec}
Another strength of the Kronecker-separable covariance model is the ability to estimate each covariance component separately, thereby reducing the overall computational complexity.
Each covariance component's sparsity enables us to efficiently estimate the clutter second-order statistics using subspace methods, such as \ac{MUSIC} \cite{stoica2005spectral}.
As it is customary in radar literature, the sensing observations are collected in the cube $\YY =\left\{[\yy_{1,v},\dots,\yy_{I,v}]\right\}_{\forall v}\in \mathds{C}^{\Mbs \times I \times V}$.
The next step would be applying a target cleaning method, such as constant false alarm rate masking \cite{rohling2007radar} or GoDec \cite{zhou2011godec}, to erase the targets from $\YY$, thereby avoiding their identification as part of the clutter.
\review{The clutter subspace here dominates the sample covariance:
this is ascribed to the point nature of the target, whereas clutter is spatially and temporally extended, occupying multiple range–Doppler bins with significantly higher power \cite{richards2010principles}. As a result, the sample covariance matrix is dominated by the clutter subspace, and the target contribution appears only as a small rank-one perturbation, to which MUSIC is known to be robust \cite{stoica2002music}.
For this reason, clutter covariance can be estimated without applying explicit target cleaning.
On the other hand, it is important to note that, in the presence of more complicated sensing scenarios, such as non-homogeneous clutter or extended targets, target cleaning is necessary. 
} 
The spatial sample covariance is computed as 
\begin{align} \label{space sample}
    \mathcal{B}_{\normalfont{\textrm{sp}}}
=\mathcal{U}_1(\overline{\YY})\mathcal{U}_1(\overline{\YY})^\hermit/IV - \sigma^2\mathbf{I}_{\Mbs} 
    \in \mathds{C}^{\Mbs \times \Mbs}.
\end{align}
The \ac{MUSIC} pseudo-spectrum of the $s$-th stream is then computed as  
$P_\textrm{sp}(\psi)=1/\mathbf{a}(\psi)^\hermit\boldsymbol{\Gamma}_{\normalfont{\textrm{sp}}}\boldsymbol{\Gamma}_{\normalfont{\textrm{sp}}}^\hermit\mathbf{a}(\psi),$
where  $   \boldsymbol{\Gamma}_{\normalfont{\textrm{sp}}}=[\boldsymbol{\lambda}_{{\normalfont{\textrm{sp}}},1}\dots,\boldsymbol{\lambda}_{{\normalfont{\textrm{sp}}},\Mbs-L}]\in \mathds{C}^{\Mbs \times \Mbs-L}$  contains the eigenvectors associated to the $\Mbs-L$ weakest eigenvectors of $\mathcal{B}_{\normalfont{\textrm{sp}}}$, i.e., its noise subspace.
The peaks of $P(\psi)$ represent the estimated clutter directions, denoted by $\hat{\psi}_n$. By substituting this into \eqref{B_s_eq}, we obtain the estimated spatial covariance matrix,denoted as $\widehat{\BB}_{\normalfont{\textrm{sp}}}$.
We once again start the estimation of the temporal covariance matrix by computing its sample counterpart, that is 
\begin{align}\label{time sample}
\mathcal{B}_{\normalfont{\textrm{t}}}=\mathcal{U}_2(\overline{\YY})\mathcal{U}_2(\overline{\YY})^\hermit /\Mbs V- \sigma^2\mathbf{I}_I \in \mathds{C}^{I \times I}.
\end{align}
The \ac{MUSIC} pseudospectrum is computed in the same way, that is 
$P_\textrm{t}(f_\rmd)=1/\bb(f_\rmd)^\hermit\boldsymbol{\Gamma}_{\normalfont{\textrm{t}}}\boldsymbol{\Gamma}_{\normalfont{\textrm{t}}}^\hermit\mathbf{b}(f_\rmd).$
Then, the peaks of this pseudospectrum are substituted into \eqref{B_t_eq} to obtain  $\widehat{\BB}_{\normalfont{\textrm{t}}}$.
The frequency sample covariance matrix is once again defined as 
\begin{align}\label{frequency sample}
\mathcal{B}_{\normalfont{\textrm{f}}}=\mathcal{U}_3(\overline{\YY})\mathcal{U}_3(\overline{\YY})^\hermit/\Mbs I - \sigma^2\mathbf{I}_V \in \mathds{C}^{V \times V}.
\end{align}
The median range of each patch can be estimated through \ac{MUSIC}.
\review{However, the presence of the diffusive component $\widetilde{\BB}_{\normalfont{\textrm{f}}}$ hinders the subspace orthogonality needed by \ac{MUSIC}. We thus adopt a diffusion-unaware approach and later quantify the performance loss caused by  $\widetilde{\BB}_{\normalfont{\textrm{f}}}$ in the numerical results section. 
}
The frequency \ac{MUSIC} pseudo-spectrum is $ P_\textrm{f}(r)
=
 \frac{1}{\dd(2r/c)^\hermit\boldsymbol{\Gamma}_{\normalfont{\textrm{f}}}\boldsymbol{\Gamma}_{\normalfont{\textrm{f}}}^\hermit\mathbf{d}(2r/c)}. $


\subsection{Clutter-Aware Matched Filtering}
We now need to perform \ac{MF} to eliminate the influence of the transmitted symbols on the radar channel.
Let us redefine vector $\boldsymbol{\mu}_{i,v}$ defined in \eqref{y radar} as 
$\boldsymbol{\mu}_{i,v} = \boldsymbol{\Phi}_{i,v}\xx_{i,v}$. Here the matrix $\boldsymbol{\Phi}_{i,v}=[\pphi_{i,v,1},\dots,\pphi_{i,v,S}]\in \mathds{C}^{\Mbs \times S}$ represents the precoded radar channel and is defined as 
\begin{align}
    \boldsymbol{\Phi}_{i,v}=[\GG_{i,v}\ff_{i,v,1}\sqrt{\rho_{i,v,1}},\dots,\GG_{i,v}\ff_{i,v,S}\sqrt{\rho_{i,v,S}}].
\end{align}
By looking at the structure of $\boldsymbol{\mu}_{i,v}$, we see that \ac{MF} can be readily applied by right-multiplying $\yy_{i,v}$ with $\xx_{i,v}^\dagger$; however, this does not suppress the clutter action.
Thanks to its low-rank structure, the clutter influence can be suppressed through whitening \cite{izquierdo2000signal,sun2017kronecker}. More specifically, the data cube $\YY$ and the symbol cube 
$\XX=\left\{ \left[\xx_{1,v},\dots,\xx_{I,v}\right] \right\}_{\forall v}$ are whitened along their time and frequency dimensions: the space dimension is left unwhitened to prevent the accidental target cancellation.
The whitening matrices are the inverse square root of the noise-adjusted covariance matrices, defined as 
\begin{align}
    &\ddt{\BB}_t^{-\frac{1}{2}}=\UU_{\normalfont{\textrm{t}}}(\boldsymbol{\Sigma}_{\normalfont{\textrm{t}}} + \sigma^2\mathbf{I}_{I}   )^{-\frac{1}{2}}\UU_{\normalfont{\textrm{t}}}^\hermit,\\
    &\ddt{\BB}_{\normalfont{\textrm{f}}}^{-\frac{1}{2}}=\UU_{\normalfont{\textrm{f}}}(\boldsymbol{\Sigma}_{\normalfont{\textrm{f}}} + \sigma^2\mathbf{I}_{V}   )^{-\frac{1}{2}}\UU_{\normalfont{\textrm{f}}}^\hermit,
\end{align} where $\widehat{\UU}_\textrm{t},\widehat{\boldsymbol{\Sigma}}_\textrm{t} (\widehat{\UU}_{\normalfont{\textrm{t}}},\widehat{\boldsymbol{\Sigma}}_{\normalfont{\textrm{f}}} )$ denote the eigen decomposition of $\sqrt[3]{\delta_\textrm{cl}^2}\widehat{\BB}_{\normalfont{\textrm{t}}}(\sqrt[3]{\delta_\textrm{cl}^2}\widehat{\BB}_{\normalfont{\textrm{f}}})$.
As it was previously mentioned, the Kronecker-separable structure allows for applying whitening independently along each dimension, that is
 \begin{align}\label{whitenign proc}
&\YY'=\mathcal{F}_2\left(\mathcal{U}_2(\YY)\ddt{\BB}_{\normalfont{\textrm{t}}}^{-\frac{1}{2}}\right),\quad\YY''=\mathcal{F}_3\left(\mathcal{U}_3(\YY')\ddt{\BB}_{\normalfont{\textrm{f}}}^{-\frac{1}{2}}\right).
 \end{align}
 After applying the same procedure to $\XX$, thus obtaining $\XX''$, the precoded radar channel is estimated as 
$\widehat{\boldsymbol{\Phi}}_{i,v}=\left[\widehat{\pphi}_{i,v,1},\dots,\widehat{\pphi}_{i,v,S} \right]= \left\{\YY'' \right\}_{i,v} \left\{\XX'' \right\}_{i,v}^\dagger. $
 \vspace{-4mm}
\subsection{Range-Angle and Range-Velocity Maps}\label{maps sec}

\reviewAfter{Once the clutter has been suppressed, we compute the \ac{RA} and \ac{RV} maps. Their peaks offer an indication of the target's angle, range, and velocity, while an explicit estimation step is not performed in this work.}
We compute the range profile by applying a \ac{DFT} along the frequency dimension on the estimated radar channels.
The range profile of the $v'$-th range bin is then computed as 
\begin{align}\label{range profile}
    \rr_{i,v',s}=\sum_{v=1}^V e^{-\frac{j2\pi v v'}{V'}} \widehat{\pphi}_{i,v,s} \in \mathds{C}^{\Mbs},
\end{align}
where $v'\in [1, V']$ and $V'\geq V$, accounting for zero padding. The delay of each target is quantized with a resolution of $1/\Delta_fV'$.
The angle profiles associated with each range profile are then computed through \ac{MUSIC}. We thus compute the sample covariance matrix of each bin's range profile as
\begin{align}\label{range cov}
    \QQ_{v',s}= \frac{1}{I}\sum_{i=1}^I \rr_{i,v',s}\rr_{i,v',s}^\hermit  \in \mathds{C}^{\Mbs \times \Mbs}.
\end{align} 
Let us denote with $\boldsymbol{\Gamma}_{v',s}$ the noise subspace of $\QQ_{v',s}$, then the angular profile of the $v'$-th bin is $P_{s,v'}(\theta)=1/{\mathbf{a}(\theta)^\hermit\boldsymbol{\Gamma}_{v',s}\boldsymbol{\Gamma}_{v',s}^\hermit\mathbf{a}(\theta)}$.
The next step is to compute the velocity profile associated with each range bin. We first project each range profile onto the angle with the highest \ac{MUSIC} power within that range bin, denoted by  $\theta^\textrm{M}_{v'}$.
Then, an inverse DFT is applied to each column of this matrix
\begin{align}\label{velocity profile}
    q_{i',v',s}=\sum_{i=1}^I e^{\frac{j2\pi i i'}{I'}}  \atx(\theta^\textrm{M}_{v'} )^\hermit\rr_{i,v',s}.
\end{align}
By collecting all the $ q_{i',v',s}$ into a matrix, the \ac{RV} map gets associated with the $s$-th stream $\VV_s \in \mathds{C}^{I' \times V'}$, where $I'\geq I$ accounts for zero padding.

\begin{table}[t]\label{table sim}
\begin{center}
\caption {\centering Simulation parameters default values}
\begin{tabular}{|m{0.65\linewidth} | m{0.25\linewidth}|} 
 \hline
 \textbf{Parameter} & \textbf{value}\\
 \hline\hline 
 BS antennas ($M_\textrm{BS})$ & $32$\\
 \hline
UE antennas ($M_\textrm{UE})$ & $2$\\
 \hline
 Active subcarriers ($V$) & $1000$\\
 \hline
 Number of transmission streams ($S$) & $3$\\
 \hline
Pilot sequence length ($\tau_\rmp$) & $3$\\
\hline
  Subcarrier spacing ($\Delta_f$) & $20$ kHz\\
 \hline
 N$^\circ$ of coherent subcarriers ($V_\textrm{cho}$) & 20 \\
 \hline
 Cyclic prefix duration ($T_\textrm{cp})$ & $3~\mu$s\\
 \hline
  N$^\circ$ of sensing \ac{OFDM} symbols($I$)  & $1000$\\
 \hline
 Carrier frequency ($f_c$) & $2$ GHz\\
 \hline
 UE's clusters angular spread ($\varsigma_{k}$) &$1^\circ$\\
 \hline
 UE's SNR per subcarrier $\left( \textrm{SNR}_v = \frac{\alpha_k^2P_\textrm{comm}}{V\sigma_k^2} \right)$ & $7+10\log_{10}(\gamma)$ dB\\
 \hline
 Clutter patches angular spread ($\varsigma_\textrm{sp} $)  & $2^\circ$\\
 \hline
  Clutter patches delay spread ($\varsigma_\tau $)  & $1$ ns\\
 \hline
 N$^\circ$ of time-coherent clutter observations ($I_\textrm{c}$) & $1000$ \\
\hline
\vspace{0.5mm}
$\widetilde{\BB}_\textrm{f}$ frequency correlation decay base ($\mu_f$)  & 0.9 \\
\hline
\vspace{0.5mm}$\widetilde{\BB}_\textrm{f}$ coherence bandwidth ($B_\textrm{cho}$) & $50$ kHz\\
\hline
\vspace{0.5mm}
$\widetilde{\BB}_\textrm{f}$ power coefficient ($\chi$) & $0$ dB\\
\hline
 UE/radar noise's powers ($\sigma^2_\ue/\sigma^2$) & $-160$ dB.\\
 \hline
 Total power budget ($P_\tx$) & $32$ dBm\\
 \hline
 Power trade-off coefficient ($\gamma$) & 0.5\\
 \hline
  Targets' RCSs powers ($\delta^2_{\textrm{Tg},l}$) & $5,1$ dBsm\\
 \hline
  Clutter's texture ($\delta^2_{\textrm{cl}}$) & $-133$ dB\\
  \hline
  Zero-padded version of $I,~V$ ($I',~V'$) & $3000,~3000$\\
  \hline
 \end{tabular}
 \end{center}
\end{table}

\section{Precoding Strategies}\label{sec prec}
We now present the precoding strategy adopted by the system.
We assume that all the subcarriers within the same coherence block have the same precoding matrix. We'll thus drop the subcarrier index from the subsequent derivations and derive the precoding matrix of an arbitrary block.
During pilot transmission, the system does not have \ac{CSIT}, or rather assumes that $\widehat{\hh}_{i}$ at the end of the previous frame has drifted too far from its actual value.
For this reason, during pilot slots, the BS relies solely on the second-order statistics of $\hh_{i}$, enabling eigen-beamforming \cite{wajid2009robust}.The $n$-th largest eigenvalue of $\CC_{\bs}$ shall be denoted by $\uu_{\bs,n}$.
When the \ac{UE} feeds back the channel estimates $\widehat{\HH}_{i}$ to the transmitter \ac{CSIT} becomes available: the \ac{BS} then employs \ac{MMSE} precoding \cite[Ch.~5]{demir2021foundations}.
This choice is motivated by the former robustness to \ac{CSI} imperfections and by the adoption of an \ac{MMSE} channel estimator.
Let us collect all the pilot slot indices in the set $\mathcal{P}$, then the precoding matrix for \ac{UE} $k$ is defined as 
\reviewAfter{
\vspace*{-2mm}
\begin{align}
&\FF_{i}^\ue=
\begin{cases}
     [\uu_{\bs,1},\dots,\uu_{\bs,\Mue}]&\textrm{if}~i\in \mathcal{P},\\
\FF_{i}^\textrm{MMSE}\triangleq\TT_{i}^{-1}\alpha_\ue\widehat{\HH}_{i}^{\top}&\textrm{otherwise},
\end{cases}
\end{align}}
where $  \TT_{i}\hspace{-1mm}=  \alpha_\ue^2(\widehat{\HH}_{i}^{\top}\widehat{\HH}_{i}^{*} + \overline{\widetilde{\XXi}}_{i})+\sigma^2_\ue\mathbf{I}_{\Mbs}$.
Here $\overline{\widetilde{\XXi}}_{i}\in \mathds{C}^{\Mbs \times \Mbs}$ is obtained by summing all the $\Mbs \times \Mbs$ blocks along the main diagonal of $\widetilde{\XXi}_{i}$.
\reviewAfter{
In the numerical results, we also use an alternative version of $\FF_{i}^\textrm{MMSE}$. The MMSE precoder can be written through its QR decomposition as   $\FF_{i}^\textrm{MMSE}=\boldsymbol{Q}_i\boldsymbol{R}_i$. We then define the QR-normalized precoder as 
$\FF_{i}^\textrm{QR}=\boldsymbol{Q}_i$.  
This preserves the subspace spanned by the original MMSE precoder but forces its columns to be orthonormal. This improves the conditioning of the communication-stream radar illumination. However, it does not make the communication precoder sensing-oriented, as the radar illumination direction is still dictated by the UE channel geometry. 
}
The sensing beam definition is independent of whether $i$ is a data or pilot slot: we adopt a beam-sweeping strategy, discretizing the sensing interval $[\theta_\textrm{start}, \theta_\textrm{end} ]$ into $\{\theta_i\}_{i=1}^I$.
\review{To ensure that the addition of a dedicated sensing stream does not disrupt the communication functions of the system,
the interference between the former and the communication streams must be eliminated. To this end, if some degree of  \ac{CSIT} is available, the sensing beams are projected onto the \ac{NS} of the communication channel estimate, thus defining the sensing precoding vector as }
\begin{align}
&\ff_{i,v,S}=
\begin{cases}
   \mathbf{a}(\theta_i)&\textrm{if}~i\in \mathcal{P},\\
    \left( \mathbf{I}_\Mbs -  \widehat{\HH}_i^\dagger\widehat{\HH}_i\right)\mathbf{a}(\theta_i)&\textrm{otherwise},\\
\end{cases}
\end{align}

\begin{figure}[t!]
\begin{center}
   \resizebox{0.48\textwidth}{!}{
    \includegraphics[scale=1]{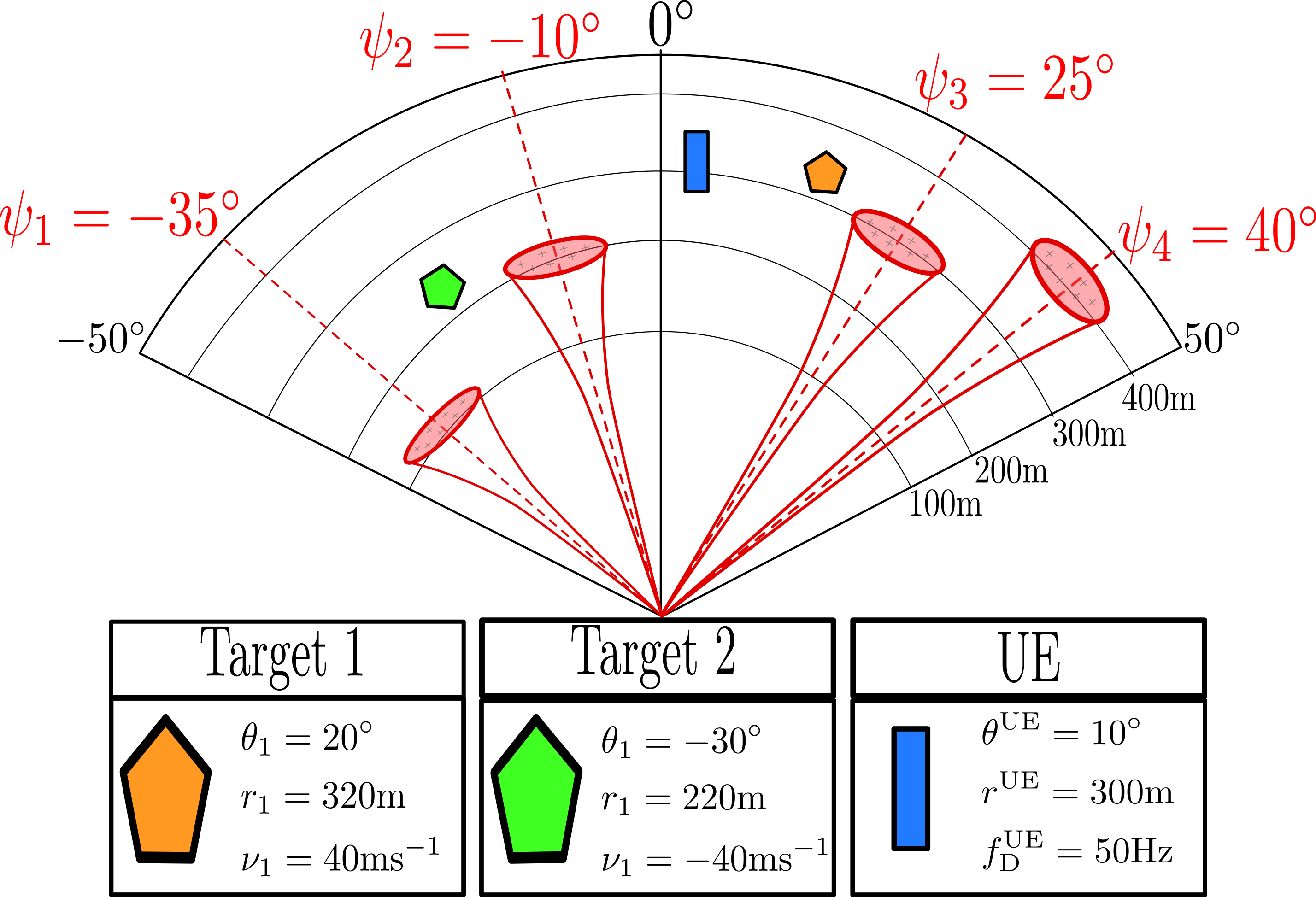}}
	 \caption{Simulated  scenario with $L=2$ radar targets, $N=4$ clutter patches and a single communication UE.}
    \label{polar radar}
		\end{center}
  \vspace{-4mm}
\end{figure}

\begin{figure}[t!]
\begin{center}
   \resizebox{0.45\textwidth}{!}{
    \includegraphics[scale=0.2]{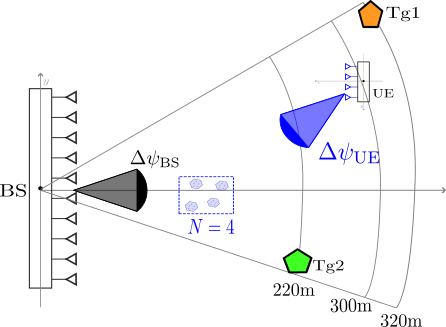}}
      \vspace{-2mm}
	 \caption{Geometric layout of the BS-UE communication channel. The set of clusters that constitute the multipath fading of $\hh_k$ does not constitute a clutter patch, as their RCSs are assumed to be too low, and thus they contribute only to $\CC_k$. Unless otherwise specified, $\Delta \psi_\textrm{BS}=[-2.5^\circ,2.5^\circ]$ and $\Delta \psi_\textrm{UE}=[-17.5^\circ,-12.5^\circ]$ }
    \label{polar UE}
		\end{center}
  \vspace{-4mm}
\end{figure}

\begin{figure}[t!]
\begin{center}
   \resizebox{0.48\textwidth}{!}{
    \includegraphics[scale=0.9]{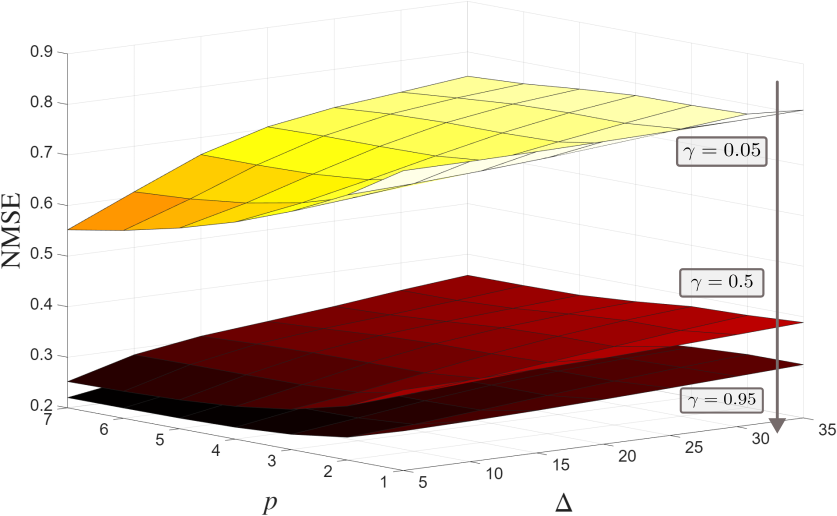}}
      \vspace{-2mm}
	 \caption{3D plot of the channel estimation NMSE performance vs the pilot spacing $\Delta$ and the number of previous pilot observations $p$ used in the channel estimation. Each surface has been obtained with a different $\gamma$ power coefficient, highlighting the trade-off between sensing and communication. }
    \label{NMSE gamma}
		\end{center}
  \vspace{-5mm}
\end{figure}
\begin{figure}[t!]
\begin{center}
   \resizebox{0.48\textwidth}{!}{
    \includegraphics[height=0.7in]{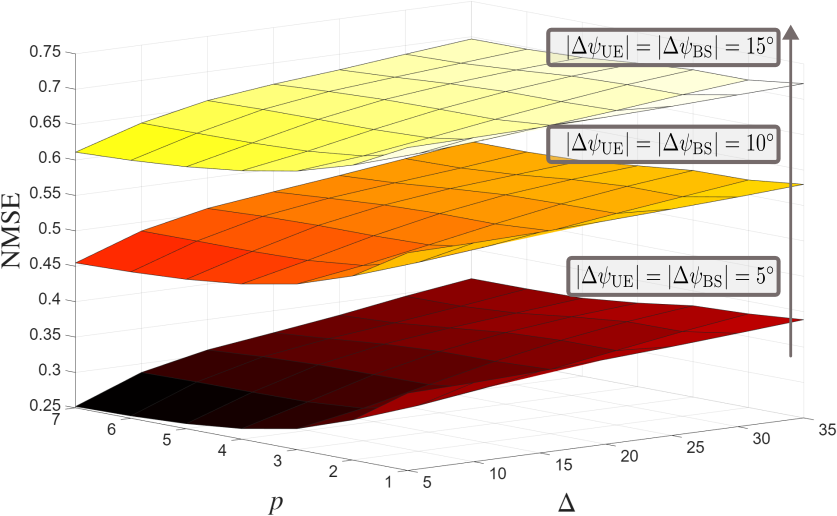}}
      \vspace{-3mm}
	 \caption{Now the same NMSE performance surfaces are computed for different values of $|\Delta\psi_{\ue}|$. We can see that the distance between clusters is directly proportional to the NMSE. }
    \label{NMSE deltapsi}
		\end{center}
  \vspace{-4mm}
\end{figure}

\begin{figure}[t!]
\begin{center}
   \resizebox{0.48\textwidth}{!}{
    \includegraphics[scale=0.9]{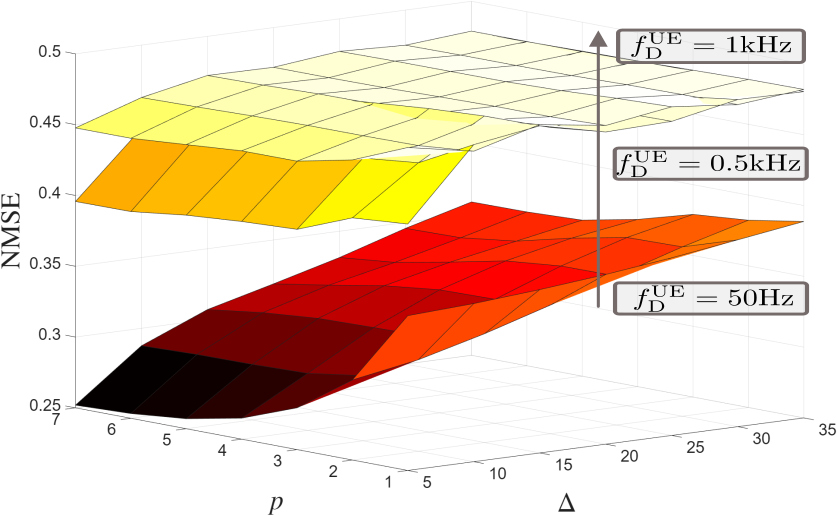}}
      \vspace{-3mm}
	 \caption{NMSE surfaces computed for different values of the UE Doppler shift $f_{\rmd}^\textrm{UE}$. As the UE's mobility increases, the surfaces become progressively flatter due to the rapid decorrelation between pilot slots.}
    \label{NMSE dopp}
		\end{center}
  \vspace{-4mm}
\end{figure}

\begin{figure}[t!]
\centering
    \begin{minipage}{\textwidth}
    \resizebox{0.48\textwidth}{!}{
    \input{images/results/communication/NMSE_along_35_block}}
    \end{minipage}
\vspace{2mm}
   \begin{minipage}{\textwidth}
    \resizebox{0.48\textwidth}{!}{
%
%
\definecolor{mycolor1}{rgb}{0.06600,0.44300,0.74500}%
\definecolor{mycolor2}{rgb}{0.86600,0.32900,0.00000}%
\definecolor{mycolor3}{rgb}{0.92900,0.69400,0.12500}%
\definecolor{mycolor4}{rgb}{0.52100,0.08600,0.81900}%
\definecolor{mycolor5}{rgb}{0.23100,0.66600,0.19600}%
\definecolor{mycolor6}{rgb}{0.18400,0.74500,0.93700}%
\definecolor{mycolor7}{rgb}{0.81900,0.01500,0.54500}%
\definecolor{mycolor8}{rgb}{0.12941,0.12941,0.12941}%
\begin{tikzpicture}

\begin{axis}[%
width=0.5\textwidth,
height=1.5in,
at={(1.888in,0.1962in)},
scale only axis,
clip=false,
xmin=0,
xmax=9,
xlabel={$i$},
ymin=0.28,
ymax=0.75,
ylabel={NMSE},
axis background/.style={fill=white},
axis x line*=bottom,
axis y line*=left,
legend columns=1,
xmajorgrids,
ymajorgrids,
legend style={at={(0.28,0.98)},legend cell align=left, align=left,font=\scriptsize}
]

\addlegendimage{ mycolor3, line width=1pt}
\addlegendentry{$f_\textrm{D}^\textrm{UE}=500$\,Hz}

\addlegendimage{ mycolor4, line width=1pt}
\addlegendentry{$f_\textrm{D}^\textrm{UE}=100$\,Hz}

\addlegendimage{ mycolor1, line width=1pt}
\addlegendentry{$f_\textrm{D}^\textrm{UE}=50$\,Hz}

\addlegendimage{ mycolor2, line width=1pt}
\addlegendentry{Block fading}

 \node[
    draw=black,
    fill=white,
    font=\scriptsize,
    anchor=north west,
    align=left,
    inner sep=4pt
] at (rel axis cs:0.28,0.9) {
\raisebox{0.3ex}{\tikz \draw[thick,solid,mark=x,mark indices={2},mark options={solid,mark size=3pt}] plot coordinates {(0,0) (0.25,0) (0.5,0)};}~$p=0$.\\
\raisebox{0.3ex}{\tikz \draw[thick,dashed, mark=square,mark indices={2}, mark options={solid}] plot coordinates {(0,0) (0.25,0) (0.5,0)};}~$p=2$\\
\raisebox{0.3ex}{\tikz \draw[thick,solid,mark=o,mark indices={2}] plot coordinates {(0,0) (0.25,0) (0.5,0)};}~$p=6$
};


\addplot [color=mycolor1, line width=1.0pt, mark=o, mark options={solid, mycolor1,mark size=3pt}, forget plot]
  table[row sep=crcr]{%
0	0.288893255505121\\
1	0.292391705717568\\
2	0.295976544592532\\
3	0.299646263419251\\
4	0.303399321111336\\
5	0.307234145116669\\
6	0.311149132344506\\
7	0.31514265010916\\
8	0.319213037089606\\
9	0.323358604304335\\
};
\addplot [color=mycolor4, line width=1.0pt, mark=square, mark options={solid, mycolor4,mark size=3pt}, forget plot]
  table[row sep=crcr]{%
0	0.327281787695346\\
1	0.33256459552161\\
2	0.338038204034042\\
3	0.343702705381985\\
4	0.349557858981177\\
5	0.355603076402658\\
6	0.361837417171341\\
7	0.36825953671208\\
8	0.374867734372796\\
9	0.381659903398251\\
};
\addplot [color=mycolor3, line width=1.0pt, mark=triangle, mark options={solid, mycolor3,mark size=3pt}, forget plot]
  table[row sep=crcr]{%
0	0.443042658233554\\
1	0.458716388962463\\
2	0.478520160279274\\
3	0.502425384361981\\
4	0.530177753033241\\
5	0.561295494852496\\
6	0.595084538964412\\
7	0.630670343111908\\
8	0.667044699686995\\
9	0.703124476718264\\
};

\addplot [color=mycolor1, dashed, line width=1.0pt, mark=square, mark options={solid, mycolor1,mark size=3pt}, forget plot]
  table[row sep=crcr]{%
0	0.314621599987346\\
1	0.316672107098279\\
2	0.318895428446649\\
3	0.321290215987063\\
4	0.323855018683767\\
5	0.326588283611216\\
6	0.329488357131443\\
7	0.332553486147261\\
8	0.335781819430241\\
9	0.339171409022341\\
};
\addplot [color=mycolor4, dashed, line width=1.0pt, mark=square, mark options={solid, mycolor4,mark size=3pt}, forget plot]
  table[row sep=crcr]{%
0	0.343431114339812\\
1	0.349978022091427\\
2	0.357029534887379\\
3	0.364570102087332\\
4	0.372583134933725\\
5	0.381051064246385\\
6	0.38995539080611\\
7	0.399276738296027\\
8	0.408994908629108\\
9	0.419088939482773\\
};
\addplot [color=mycolor3, dashed, line width=1.0pt, mark=square, mark options={solid, mycolor3,mark size=3pt}, forget plot]
  table[row sep=crcr]{%
0	0.453735327080593\\
1	0.472206003990065\\
2	0.495867929336242\\
3	0.524332711134092\\
4	0.55693309595151\\
5	0.592764022821827\\
6	0.630739815762661\\
7	0.669663095526974\\
8	0.708300204859973\\
9	0.74545756928622\\
};

\addplot [black,line width=1.3pt,forget plot, mark=x, mark options={solid, black, mark size=4.2pt}, forget plot]
  table[row sep=crcr]{%
0	0.4822\\
1	0.4822\\
2	0.4822\\
3	0.4822\\
4	0.4822\\
5	0.4822\\
6	0.4822\\
7	0.4822\\
8	0.4822\\
9	0.4822\\
};
\addplot [mycolor2,line width=1pt,forget plot, mark=x, mark options={solid, mycolor2, mark size=4pt}, forget plot]
  table[row sep=crcr]{%
0	0.4822\\
1	0.4822\\
2	0.4822\\
3	0.4822\\
4	0.4822\\
5	0.4822\\
6	0.4822\\
7	0.4822\\
8	0.4822\\
9	0.4822\\
};








\node at (axis description cs:0.5,-0.28) {(b)};

\end{axis}

\end{tikzpicture}%
}
    \end{minipage}
\vspace{-3mm}
    \caption{\review{NMSE progression within a frame of a) $\Delta=35$ and b) data slots $\Delta=10$. We see that the UE mobility has a much bigger impact than $p$, and the impact of $p$ is bigger with smaller frames.    }}
\vspace{-3mm}
\label{subfig along}
\end{figure}


\begin{figure}[t!]
\begin{center}
   \resizebox{0.48\textwidth}{!}{
%
%
\usetikzlibrary{arrows.meta}
\definecolor{mycolor1}{rgb}{0.06600,0.44300,0.74500}%
\definecolor{mycolor2}{rgb}{0.86600,0.32900,0.00000}%
\definecolor{mycolor3}{rgb}{0.92900,0.69400,0.12500}%
\definecolor{mycolor4}{rgb}{0.52100,0.08600,0.81900}%
\definecolor{mycolor5}{rgb}{0.23100,0.66600,0.19600}%
\definecolor{mycolor6}{rgb}{0.18400,0.74500,0.93700}%
\definecolor{mycolor7}{rgb}{0.81900,0.01500,0.54500}%
\definecolor{mycolor8}{rgb}{0.12941,0.12941,0.12941}%
\begin{tikzpicture}

\begin{axis}[%
width=0.5\textwidth,
height=1.5in,
at={(1.888in,0.962in)},
scale only axis,
xmin=0,
xmax=9,
xlabel={$i$},
ymin=0.25,
ymax=0.75,
ylabel={NMSE},
axis background/.style={fill=white},
axis x line*=bottom,
axis y line*=left,
legend columns=1,
xmajorgrids,
ymajorgrids,
set layers,
legend style={at={(0.3,0.75)},legend cell align=left, align=left,font=\scriptsize}
]

\addlegendimage{ mycolor7, line width=1pt}
\addlegendentry{$\textrm{SNR}_v=-5$\,dB}

\addlegendimage{ mycolor6, line width=1pt}
\addlegendentry{$\textrm{SNR}_v=0$\,dB}

\addlegendimage{ mycolor5, line width=1pt}
\addlegendentry{$\textrm{SNR}_v=4$\,dB}

 \node[
    draw=black,
    fill=white,
    font=\scriptsize,
    anchor=north west,
    align=left,
    inner sep=4pt
] at (rel axis cs:0.8,0.7) {
\raisebox{0.3ex}{\tikz \draw[thick,dashed, mark=square,mark indices={2}, mark options={solid}] plot coordinates {(0,0) (0.25,0) (0.5,0)};}~$p=2$\\
\raisebox{0.3ex}{\tikz \draw[thick,solid,mark=o,mark indices={2}] plot coordinates {(0,0) (0.25,0) (0.5,0)};}~$p=6$
};


\addplot [color=mycolor5, line width=1.0pt, mark=o, mark options={solid, mycolor5,mark size=2.5pt},forget plot]
  table[row sep=crcr]{%
0	0.27438975460221\\
1	0.277173591153656\\
2	0.280029870774057\\
3	0.282957689118902\\
4	0.285956120490976\\
5	0.289024218237743\\
6	0.292161015157357\\
7	0.295365523913135\\
8	0.298636737456264\\
9	0.30197362945653\\
};

\addplot [color=mycolor6, line width=1.0pt, mark=o, mark options={solid, mycolor6,mark size=2.5pt},forget plot]
  table[row sep=crcr]{%
0	0.368195541336966\\
1	0.371216382574204\\
2	0.374314401946666\\
3	0.377488497531511\\
4	0.380737541845577\\
5	0.384060382366635\\
6	0.387455842065825\\
7	0.390922719950967\\
8	0.394459791620474\\
9	0.398065809827586\\
};

\addplot [color=mycolor7, line width=1.0pt, mark=o, mark options={solid, mycolor7,mark size=2.5pt},forget plot]
  table[row sep=crcr]{%
0	0.601230177064066\\
1	0.603405912228862\\
2	0.605636712886233\\
3	0.607921741698178\\
4	0.610260142052399\\
5	0.612651038470364\\
6	0.615093537024065\\
7	0.617586725761254\\
8	0.62012967513889\\
9	0.62272143846462\\
};

\addplot [color=mycolor5, line width=1.0pt, mark=square, dashed, mark options={solid, mycolor5,mark size=2.5pt},forget plot]
  table[row sep=crcr]{%
0	0.310396656743083\\
1	0.311747513354092\\
2	0.313213491980826\\
3	0.314794017817038\\
4	0.31648847140361\\
5	0.318296188930942\\
6	0.32021646256292\\
7	0.322248540782271\\
8	0.324391628757132\\
9	0.32664488872863\\
};

\addplot [color=mycolor6, line width=1.0pt, mark=square,dashed, mark options={solid, mycolor6,mark size=2.5pt},forget plot]
  table[row sep=crcr]{%
0	0.45456892516651\\
1	0.455658831721459\\
2	0.456841642228047\\
3	0.458116889636531\\
4	0.459484070606556\\
5	0.460942645753747\\
6	0.462492039913891\\
7	0.464131642424579\\
8	0.465860807424136\\
9	0.467678854167696\\
};

\addplot [color=mycolor7, line width=1.0pt, mark=square,dashed, mark options={solid, mycolor7,mark size=2.5pt},forget plot]
  table[row sep=crcr]{%
0	0.731223856985216\\
1	0.731766367157966\\
2	0.732355125262172\\
3	0.732989898008312\\
4	0.733670433978045\\
5	0.734396463747594\\
6	0.735167700019947\\
7	0.735983837765783\\
8	0.736844554373051\\
9	0.73774950980513\\
};

\begin{pgfonlayer}{axis foreground}
\draw[
  ultra thick,
  draw=black,
  {Latex[length=8pt,width=8pt]}-{Latex[length=8pt,width=8pt]}
]
  (axis cs:4,0.610260142052399) -- (axis cs:4,0.733670433978045);
\draw[
  thick,
  draw=mycolor2,
  {Latex[length=5.5pt,width=5.5pt]}-{Latex[length=5.5pt,width=5.5pt]}
]
  (axis cs:4,0.618260142052399) -- (axis cs:4,0.725670433978045);

\draw[
  ultra thick,
  draw=black,
  {Latex[length=5pt,width=5pt]}-{Latex[length=5pt,width=5pt]}
]
  (axis cs:4,0.380737541845577) -- (axis cs:4,0.459484070606556);
\draw[
  thick,
  draw=mycolor2,
  {Latex[length=3pt,width=3pt]}-{Latex[length=3pt,width=3pt]}
]
  (axis cs:4,0.38737541845577) -- (axis cs:4,0.453484070606556);

\draw[
  very thick,
  draw=black,
  {Latex[length=3pt,width=3pt]}-{Latex[length=3pt,width=3pt]}
]
  (axis cs:4,0.285956120490976) -- (axis cs:4,0.31648847140361);
\draw[
  thick,
  draw=mycolor2,
  {Latex[length=2pt,width=2pt]}-{Latex[length=2pt,width=2pt]}
]
  (axis cs:4,0.28956120490976) -- (axis cs:4,0.31248847140361);

\end{pgfonlayer}

\end{axis}
\end{tikzpicture}%
}
      \vspace{-3mm}
	 \caption{
     \review{NMSE progression along the frame with $\Delta=10$ and $f_\textrm{D}^\textrm{UE}=50$\,Hz for different values of $\textrm{SNR}_v$. We see that the benefit of using $p>1$ previous pilot observation is inversely proportional to $\textrm{SNR}_v$ }
     }
    \label{gap vs p}
		\end{center}
  \vspace{-5mm}
\end{figure}

\section{Numerical Results}\label{num res}
\reviewAfter{The effects of channel aging on channel estimation and the proposed clutter estimation and suppression procedure are numerically evaluated in the scenario depicted in Figs.~\ref {polar radar} and \ref{polar UE}.}
The said figures also report the default values of the simulation parameters, the rest of which are listed in Table II.
We assume that the BS employs equal power allocation between streams and subcarriers, that is $\boldsymbol{\rho}_{i,v}=\left[\sqrt{\frac{P_\textrm{comm}}{ M_\textrm{UE}V} }, \dots,\sqrt{\frac{P_\textrm{comm}}{ M_\textrm{UE}V} }, \sqrt{\frac{P_\textrm{sens}}{ V} }    \right]\in \mathbb{C}^{S}, \forall i,v$.
\reviewAfter{Unless otherwise specified, the precoding matrix used during the data slot is $\FF_i^\textrm{MMSE}$.
An additional quantitative metric used to assess the proposed clutter estimation and suppression procedure is the targets' \ac{TFR}, namely the ratio between the target's peak power and the residual clutter floor. This metric is an indication of the target identifiability in the generated maps \cite{kim2019improvement}. }
Let us denote the RA map of stream $s$ expressed in dB scale as $\mathcal{M}_s(\theta,r)$. Let us further denote with $\hat{r}_l,\hat{\theta}_l$ the closest range and angle bins to the respective ground truths: the \ac{TFR} of the $s$-th stream map for the $l$-th target is then computed as 
\begin{align}
\textrm{TFR}_s=\mathcal{M}_s(\hat{\theta}_l,\hat{r}_l) - \textrm{median}\left[ \widetilde{\mathcal{M}_s}(\theta,r)\right]~\textrm{[dB]}
\end{align}
where $\widetilde{\mathcal{M}}_s(\theta,r)= \mathcal{M}_s(\theta,r)- \cup_l \mathcal{G}_l$, with $\mathcal{G}_l$ denoting a guard region around each target's peak. 
Unless otherwise specified, the following numerical results consider unperturbed clutter.
\begin{figure}[t!]
\begin{center}
   \resizebox{0.48\textwidth}{!}{
%
%
\definecolor{mycolor1}{rgb}{0.06600,0.44300,0.74500}%
\definecolor{mycolor2}{rgb}{0.86600,0.32900,0.00000}%
\definecolor{mycolor3}{rgb}{0.92900,0.69400,0.12500}%
\definecolor{mycolor4}{rgb}{0.52100,0.08600,0.81900}%
\definecolor{mycolor5}{rgb}{0.12941,0.12941,0.12941}%
\begin{tikzpicture}

\begin{axis}[%
width=0.5\textwidth,
height=1.3in,
at={(1.888in,0.962in)},
scale only axis,
xmin=0,
xmax=500,
xlabel style={font=\color{mycolor5}},
xlabel={$B_\textrm{cho}$[kHz]},
ymin=-35,
ymax=0,
ylabel style={font=\color{mycolor5}},
ylabel={NMSE},
axis background/.style={fill=white},
axis x line*=bottom,
axis y line*=left,
legend columns=2,
xmajorgrids,
ymajorgrids,
legend style={at={(1,0.3)},legend cell align=left, align=left, font=\scriptsize}
]
\addplot [color=mycolor1, line width=1.0pt, mark=o, mark options={solid, mycolor1,mark size=3pt}]
  table[row sep=crcr]{%
5	-20.2233376612703\\
75.7142857142857	-18.7929905075293\\
146.428571428571	-18.1498179207267\\
217.142857142857	-17.6437879128285\\
287.857142857143	-17.3827076562861\\
358.571428571429	-17.1941370255377\\
429.285714285714	-17.023669090328\\
500	-16.9087659702601\\
};
\addlegendentry{space}

\addplot [color=mycolor2, line width=1.0pt, mark=square, mark options={solid, mycolor2,mark size=3pt}]
  table[row sep=crcr]{%
5	-21.2475812213244\\
75.7142857142857	-21.5703122082477\\
146.428571428571	-21.4600488911345\\
217.142857142857	-21.0850358348691\\
287.857142857143	-21.0850358348691\\
358.571428571429	-21.0850358348691\\
429.285714285714	-11.0172990091218\\
500	-8.24303357885043\\
};
\addlegendentry{time}

\addplot [color=mycolor3, line width=1.0pt, mark=triangle, mark options={solid, mycolor3,mark size=3pt}]
  table[row sep=crcr]{%
5	-19.9116188902705\\
75.7142857142857	-9.13117916983998\\
146.428571428571	-6.77241212079002\\
217.142857142857	-4.64183062927702\\
287.857142857143	-4.57067686036095\\
358.571428571429	-4.32243856493238\\
429.285714285714	-4.11396623766808\\
500	-3.93508280334942\\
};
\addlegendentry{frequency}

\addplot [color=mycolor4, line width=1.0pt, mark=x, mark options={solid, mycolor4,mark size=3pt}]
  table[row sep=crcr]{%
5	-32.4638233987374\\
75.7142857142857	-32.044160958535\\
146.428571428571	-32.7116652513485\\
217.142857142857	-5.22746359380073\\
287.857142857143	-6.47544325783804\\
358.571428571429	-6.4767435425004\\
429.285714285714	-6.47868975658309\\
500	-6.4784600583863\\
};
\addlegendentry{frequency sparse}

\end{axis}
\end{tikzpicture}%
}
      \vspace{-2mm}
	 \caption{Clutter estimation NMSE as a function of  $B_\textrm{cho}$}
    \label{CLNMSE_vs_Bcho}
		\end{center}
  \vspace{-4mm}
\end{figure}
  \vspace{-4mm}
\begin{figure}[t!]
\begin{center}
   \resizebox{0.48\textwidth}{!}{
%
%
\definecolor{mycolor1}{rgb}{0.06600,0.44300,0.74500}%
\definecolor{mycolor2}{rgb}{0.86600,0.32900,0.00000}%
\definecolor{mycolor3}{rgb}{0.92900,0.69400,0.12500}%
\definecolor{mycolor4}{rgb}{0.52100,0.08600,0.81900}%
\definecolor{mycolor5}{rgb}{0.12941,0.12941,0.12941}%
\begin{tikzpicture}

\begin{axis}[%
width=0.5\textwidth,
height=1.3in,
at={(1.888in,0.962in)},
scale only axis,
xmin=-10,
xmax=10,
xlabel style={font=\color{mycolor5}},
xlabel={$\chi$[dB]},
ymin=-40,
ymax=2,
ylabel style={font=\color{mycolor5}},
ylabel={NMSE},
axis background/.style={fill=white},
axis x line*=bottom,
axis y line*=left,
legend columns=2,
xmajorgrids,
ymajorgrids,
legend style={at={(1,0.26)},legend cell align=left, align=left,font=\scriptsize}
]
\addplot [color=mycolor1, line width=1.0pt, mark=o, mark options={solid, mycolor1,mark size=3pt}]
  table[row sep=crcr]{%
-10	-14.4912297321099\\
-7.14285714285717	-15.1778459797044\\
-4.28571428571428	-16.3062834336103\\
-1.42857142857144	-18.0719375011082\\
1.42857142857144	-20.5128665674204\\
4.28571428571428	-23.4778992973245\\
7.14285714285717	-25.1846541609171\\
10	-26.4846501963543\\
};
\addlegendentry{space}

\addplot [color=mycolor2, line width=1.0pt, mark=square, mark options={solid, mycolor2,mark size=3pt}]
  table[row sep=crcr]{%
-10	-9.86231425395306\\
-7.14285714285717	-9.86231425395306\\
-4.28571428571428	-9.86231425395306\\
-1.42857142857144	-21.1451223028059\\
1.42857142857144	-21.5703122082477\\
4.28571428571428	-21.5703122082477\\
7.14285714285717	-21.3227443784583\\
10	-21.3227443784583\\
};
\addlegendentry{time}

\addplot [color=mycolor3, line width=1.0pt, mark=triangle, mark options={solid, mycolor3,mark size=3pt}]
  table[row sep=crcr]{%
-10	-29.698114452836\\
-7.14285714285717	-24.3600735895585\\
-4.28571428571428	-18.8618547607192\\
-1.42857142857144	-13.3782278796887\\
1.42857142857144	-8.25815755549047\\
4.28571428571428	-3.87659421637255\\
7.14285714285717	-1.43822732819183\\
10	-0.820548540057587\\
};
\addlegendentry{frequency}

\addplot [color=mycolor4, line width=1.0pt, mark=x, mark options={solid, mycolor4,mark size=3pt}]
  table[row sep=crcr]{%
-10	-38.6459438820601\\
-7.14285714285717	-37.11365477812\\
-4.28571428571428	-36.0756273868616\\
-1.42857142857144	-33.4030305649696\\
1.42857142857144	-29.7471060837758\\
4.28571428571428	-11.0662230812752\\
7.14285714285717	-1.64542989338923\\
10	0.713914102077808\\
};
\addlegendentry{frequency sparse}

\end{axis}
\end{tikzpicture}%
}
      \vspace{-2mm}
		 \caption{Clutter estimation NMSE as a function of  $\chi$}
    \label{CLNMSE_vs_chi}
		\end{center}
  \vspace{-4mm}
\end{figure}


\begin{figure*}[h!]
\begin{center}
   \resizebox{0.95\textwidth}{!}{
    \includegraphics[scale=1]{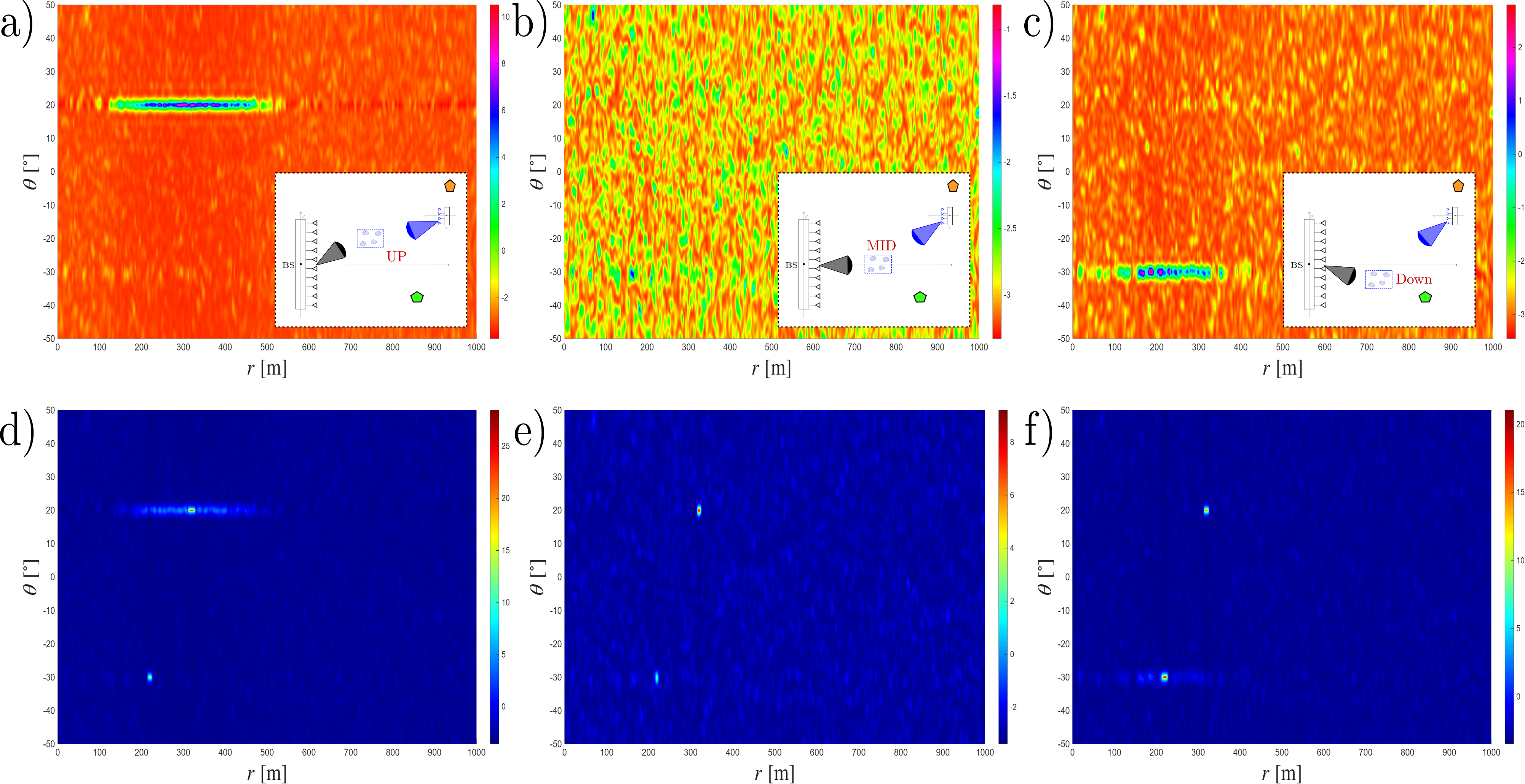}}
      \vspace{-2mm}
	 \caption{RA map in dB scale associated with $s=1$ using different data-slot precoders, namely $\FF_i^\textrm{MMSE}$(top) and $\FF_i^\textrm{QR}$ (bottom). We have subtracted $80\%$ of their first singular value from these maps to have a clearer map. This partially removes the horizontal stripe across $r$, which indicates energy leakage across range due to the high \ac{SNR} of the targets. Each target peak is at least 10 dB higher than its stripe, so this does not constitute a source of confusion.}
    \label{RA comm}
		\end{center}
  \vspace{-3mm}

\end{figure*}

\begin{figure*}[h!]
\begin{center}
   \resizebox{0.95\textwidth}{!}{
    \includegraphics[scale=1.2]{images/results/RA_maps/maps_sens_tot.png}}
      \vspace{-1mm}
	 \caption{Range-angle map in dB scale associated with  $s=S$ for different levels of clutter awareness, that is: a) \ac{MF} without whitening, b) estimated covariances, and c) \ac{MF} true covariances. We have implemented the same singular value subtraction and a Kaiser window of $3$rd order.}
    \label{RA sens}
		\end{center}
  \vspace{-3mm}
\end{figure*}

\begin{figure}[h!]
\begin{center}
   \resizebox{0.48\textwidth}{!}{
%
%
\definecolor{mycolor1}{rgb}{0.12941,0.12941,0.12941}%
\begin{tikzpicture}

\begin{axis}[%
width=0.5\textwidth,
height=1.5in,
at={(1.888in,0.809in)},
scale only axis,
xmin=10,
xmax=100,
xlabel style={font=\color{mycolor1}},
xlabel={$\gamma~[\%]$},
ymin=-5,
ymax=45,
ylabel style={font=\color{mycolor1}},
ylabel={$\textrm{TFR}_S$ [dB]},
axis background/.style={fill=white},
axis x line*=bottom,
axis y line*=left,
xmajorgrids,
ymajorgrids,
legend style={at={(0.406,0.1)}, anchor=south west, legend cell align=left, align=left}
]


\addlegendimage{ blue, line width=1pt}
\addlegendentry{Target 1}

\addlegendimage{red, line width=1pt}
\addlegendentry{Target 2}

 \node[
    draw=black,
    fill=white,
    font=\small,
    anchor=north west,
    align=left,
    inner sep=4pt
] at (rel axis cs:0.1,0.28) {
\raisebox{0.3ex}{\tikz \draw[thick,solid,mark=square,mark indices={2}] plot coordinates {(0,0) (4,0) (8,0)};}~Perfect\\
\raisebox{0.3ex}{\tikz \draw[thick,dashed, mark=triangle,mark indices={2}, mark options={solid}] plot coordinates {(0,0) (4,0) (8,0)};}~Estimated
};


\addplot [color=blue, line width=1.0pt, mark size=2pt, mark=square, mark options={solid, blue},forget plot]
  table[row sep=crcr]{%
10	40.6182065075046\\
14.7368421052632	40.4515848426728\\
19.4736842105263	40.2713536496959\\
24.2105263157895	40.0754277448976\\
28.9473684210526	39.8615913775758\\
33.6842105263158	39.6271008110229\\
38.4210526315789	39.3685076557041\\
43.1578947368421	39.0814644408056\\
47.8947368421053	38.7604146748278\\
52.6315789473684	38.3980823809251\\
57.3684210526316	37.984663247\\
62.1052631578947	37.506572279038\\
66.8421052631579	36.9443904157192\\
71.5789473684211	36.2688127890343\\
76.3157894736842	35.4329437132854\\
81.0526315789474	34.3551211293293\\
85.7894736842105	32.8749401029059\\
90.5263157894737	30.6109183017057\\
95.2631578947368	26.2438894176521\\
100	0.0557033190906783\\
};
\addlegendentry{Perfect, Tg 1}

\addplot [color=blue, dashed, line width=1.0pt, mark size=2pt, mark=triangle, mark options={solid, blue},forget plot]
  table[row sep=crcr]{%
10	19.3895909604371\\
14.7368421052632	19.3661989731629\\
19.4736842105263	19.3448331750644\\
24.2105263157895	19.3238488488206\\
28.9473684210526	19.3024585595484\\
33.6842105263158	19.2787968225688\\
38.4210526315789	19.2520511035352\\
43.1578947368421	19.2200644958007\\
47.8947368421053	19.1787813308348\\
52.6315789473684	19.1239443586838\\
57.3684210526316	19.0481470185571\\
62.1052631578947	18.9366719313798\\
66.8421052631579	18.7713059015983\\
71.5789473684211	18.5256229374354\\
76.3157894736842	18.1577225429876\\
81.0526315789474	17.5996142197955\\
85.7894736842105	16.7760886421227\\
90.5263157894737	15.3775548713547\\
95.2631578947368	11.8964379360121\\
100	0.311491715891013\\
};
\addlegendentry{Estimated, Tg 1}

\addplot [color=red, line width=1.0pt, mark size=2pt, mark=square, mark options={solid, red},forget plot]
  table[row sep=crcr]{%
10	33.8589664195205\\
14.7368421052632	33.8705518230248\\
19.4736842105263	33.8778352509616\\
24.2105263157895	33.8805876955339\\
28.9473684210526	33.8784106873576\\
33.6842105263158	33.8707275449259\\
38.4210526315789	33.8567767371798\\
43.1578947368421	33.835487119104\\
47.8947368421053	33.805405583054\\
52.6315789473684	33.7644910925081\\
57.3684210526316	33.7097829977882\\
62.1052631578947	33.6367599798935\\
66.8421052631579	33.5383607276223\\
71.5789473684211	33.402998374904\\
76.3157894736842	33.2102139155452\\
81.0526315789474	32.9200570637508\\
85.7894736842105	32.4425681143599\\
90.5263157894737	31.5248241076943\\
95.2631578947368	29.0909318175382\\
100	0.0751358602180854\\
};
\addlegendentry{Perfect, Tg 2}

\addplot [color=red, dashed, line width=1.0pt, mark size=2pt, mark=triangle, mark options={solid, red},forget plot]
  table[row sep=crcr]{%
10	24.067118603133\\
14.7368421052632	24.0978392082872\\
19.4736842105263	24.1122664676043\\
24.2105263157895	24.1066520865088\\
28.9473684210526	24.0777345862346\\
33.6842105263158	24.0223430253571\\
38.4210526315789	23.9367471187695\\
43.1578947368421	23.8165405665433\\
47.8947368421053	23.6558756771993\\
52.6315789473684	23.4472165112559\\
57.3684210526316	23.1803894815966\\
62.1052631578947	22.8480092018165\\
66.8421052631579	22.4376478483943\\
71.5789473684211	21.9281962975717\\
76.3157894736842	21.2839100753508\\
81.0526315789474	20.4385689534546\\
85.7894736842105	19.2599450111033\\
90.5263157894737	17.4359023600512\\
95.2631578947368	14.1039746429246\\
100	1.07073785943943\\
};
\addlegendentry{Estimated, Tg 2}

\end{axis}
\end{tikzpicture}%
}
      \vspace{-1mm}
	 \caption{
    \review{\ac{TFR} associated with the sensing stream of both targets for imperfect and perfect clutter knowledge. The RA maps underwent
     the same singular value subtraction of Fig. \ref{RA comm}. This figure has been obtained by averaging the TFRs of 15 Monte Carlo realizations of $\YY$.} }
    \label{TFR high gamma}
		\end{center}
  \vspace{-3mm}
\end{figure}

\begin{figure}[t!]
\begin{center}
   \resizebox{0.48\textwidth}{!}{
%
%
\definecolor{mycolor1}{rgb}{0.06600,0.44300,0.74500}%
\definecolor{mycolor2}{rgb}{0.86600,0.32900,0.00000}%
\definecolor{mycolor3}{rgb}{0.92900,0.69400,0.12500}%
\definecolor{mycolor4}{rgb}{0.52100,0.08600,0.81900}%
\definecolor{mycolor5}{rgb}{0.12941,0.12941,0.12941}%
\begin{tikzpicture}

\begin{axis}[%
width=0.5\textwidth,
height=1.3in,
at={(1.888in,0.962in)},
scale only axis,
xmin=0,
xmax=100,
xlabel style={font=\color{mycolor5}},
xlabel={$\varsigma_\tau$[ns]},
ymin=-31,
ymax=0,
ylabel style={font=\color{mycolor5}},
ylabel={NMSE},
axis background/.style={fill=white},
axis x line*=bottom,
axis y line*=left,
legend columns=2,
xmajorgrids,
ymajorgrids,
legend style={at={(1,0.52)},legend cell align=left, align=left, font=\scriptsize}
]
\addplot [color=mycolor1, line width=1.0pt, mark=o, mark options={solid, mycolor1,mark size=3pt}]
  table[row sep=crcr]{%
1	-19.0729147619157\\
15.1428571428571	-22.4801440308133\\
29.2857142857143	-24.1526991763456\\
43.4285714285714	-25.0165039046373\\
57.5714285714286	-25.069688105464\\
71.7142857142857	-25.4649022941767\\
85.8571428571428	-26.0331498486029\\
100	-26.3468984891271\\
};
\addlegendentry{space}

\addplot [color=mycolor2, line width=1.0pt, mark=square, mark options={solid, mycolor2,mark size=3pt}]
  table[row sep=crcr]{%
1	-21.5703122082477\\
15.1428571428571	-21.5703122082477\\
29.2857142857143	-21.5703122082477\\
43.4285714285714	-21.7376223346967\\
57.5714285714286	-21.7376223346967\\
71.7142857142857	-21.7376223346967\\
85.8571428571428	-21.7376223346967\\
100	-21.7376223346967\\
};
\addlegendentry{time}

\addplot [color=mycolor3, line width=1.0pt, mark=triangle, mark options={solid, mycolor3,mark size=3pt}]
  table[row sep=crcr]{%
1	-10.7446064115951\\
15.1428571428571	-8.92152582702948\\
29.2857142857143	-7.16361071715266\\
43.4285714285714	-5.76726673228933\\
57.5714285714286	-3.69161668454325\\
71.7142857142857	-2.79828151863376\\
85.8571428571428	-3.94733127277152\\
100	-3.47308321931928\\
};
\addlegendentry{frequency}

\addplot [color=mycolor4, line width=1.0pt, mark=x, mark options={solid, mycolor4,mark size=3pt}]
  table[row sep=crcr]{%
1	-31.9361949425861\\
15.1428571428571	-20.2910550459649\\
29.2857142857143	-16.9003129375199\\
43.4285714285714	-12.2958478402571\\
57.5714285714286	-5.63445530657282\\
71.7142857142857	-4.02305010407343\\
85.8571428571428	-3.18254784170892\\
100	-0.535800893076896\\
};
\addlegendentry{frequency sparse}

\end{axis}
\end{tikzpicture}%
}
      \vspace{-2mm}
	 \caption{Clutter estimation NMSE as a function of   $\varsigma_\tau$}
    \label{CLNMSE_vs_del}
		\end{center}
  \vspace{-5mm}
\end{figure}

\begin{figure}[h!]
\begin{center}
 \resizebox{0.48\textwidth}{!}{
   \input{images/results/beampattern/beamp_B_f_500khz}}
      \vspace{-2mm}
	 \caption{beampattern of $\BB_\textrm{f}$ and $\widehat{\BB}_\textrm{f}$ with $B_\textrm{cho}=500$ kHz}
    \label{beamp_bcho}
		\end{center}
  \vspace{-5mm}
\end{figure}
\begin{figure}[h!]
\begin{center}
 \resizebox{0.48\textwidth}{!}{
   \input{images/results/beampattern/beamp_B_f_chi_8}}
      \vspace{-2mm}
	 \caption{beampattern of $\BB_\textrm{f}$ and $\widehat{\BB}_\textrm{f}$ with $\chi=8$ dB}
    \label{beamp_chi}
		\end{center}
  \vspace{-5mm}
\end{figure}

\subsection{Communication Channel Estimation}
The channel $\hh_i$ is estimated during pilot slots.
This estimate depends on the frame size, $\Delta$. The channel estimate in Corollary 1 is computed using $p$ previous pilot observations, thereby leveraging the channel's known time-correlation structure. $\CC $ has been normalized to $\Tr(\CC )=1$, allowing for a cleaner \ac{SNR} expression.
The metric chosen to assess the channel estimation quality is the \ac{NMSE}, 
defined as $\textrm{NMSE}_{i} =\textrm{Tr}\left( \widetilde{\boldsymbol{\Xi}}_{i} \right)$.
We first investigate the channel estimation performance during a generic pilot slot: Fig.~\ref{NMSE gamma}-\ref{NMSE dopp} show 3D surfaces with $\Delta$ on the $x$ axis,  $p$ on the $y$ axis, and the \ac{NMSE} on the $z$ axis. Fig.~\ref{NMSE gamma} is a visual representation of the sensing-communication trade-off, as a higher $\gamma$ corresponds to a lower \ac{NMSE}. 
This hardly comes as a surprise since $\gamma$ represents the fraction of power dedicated to the communication beams. Interestingly, the performance gap between $\gamma=0.05$ and $\gamma=0.5$ is much higher than the gap between the latter and $\gamma=0.95$.
 Fig.~\ref{NMSE deltapsi} shows the impact of $\Delta\psi_{\textrm{BS}}$  on the \ac{NMSE}.
 The \ac{NMSE} degradation is proportional to $\Delta\psi_{\textrm{BS}}$, which is caused by the fact that a greater distance between clusters makes the reflection process more diffusive than specular.
Next, we analyze the \ac{NMSE} dependence on the \ac{UE}'s mobility, measured through its Doppler shift: Fig. \ref{NMSE dopp} shows that higher mobility implies a performance loss, but, more interestingly, a loss in sensitivity to $\Delta$ and $p$, as the \ac{NMSE} surfaces become progressively flatter with $f_\textrm{D}^\textrm{UE}$. This is caused by the rapid decorrelation induced by high mobility, which ultimately makes the estimator defined in Lemma \ref{lemma hk} collapse onto a standard block fading \ac{MMSE} estimator.
Indeed, a block fading estimator is obtained from Lemma \ref{lemma hk} by setting $p=0$.
Our investigation continues with  Fig. \ref{subfig along}: within the frame, the estimate drifts away from the real channel value, causing the \ac{NMSE} to increase with $i$.
It goes without saying that the block fading \ac{NMSE} does not change along the frame. These figures show exactly when our approach outperforms a block fading one. Indeed, a block fading model is the better alternative when \acp{UE} have too high a Doppler shift compared to when $\Delta$ is big.
By reducing $\Delta$ when \acp{UE} show high Doppler, our channel estimator dramatically outperforms block fading.
\review{Finally, Fig. \ref{gap vs p} shows how the performance benefit of utilizing more than 1 pilot observation is inversely proportional to $\textrm{SNR}_v$. The performance gap, highlighted by the red double arrow, is indeed big when $\textrm{SNR}_v=-5$ dB, while it progressively reduces as the $\textrm{SNR}_v$ increases.
It should be noted that, if more than one UE were present, as long as pilot orthogonality is preserved, the provided channel estimation performance trends presented here would still hold. They would only be shifted upwards on the z or y axis, due to a reduced $\textrm{SNR}_v$ per UE as the BS would need to spread its power more thinly.
Furthermore, reducing $\Delta$ would inevitably reduce the \acp{UE} \ac{SE}. We reserve a careful investigation of SE for future work, even though a trade-off between channel estimation accuracy, inversely proportional to $\Delta$, and the fraction of data slots in the frame is expected to manifest itself 
\cite{fodor2023optimizing}. }

\begin{figure}[t!]
\begin{center}
   \resizebox{0.48\textwidth}{!}{
   \input{images/results/multi-UE-loss/FIG_eta}}
      \vspace{-2mm}
	 \caption{NS projection loss vs steering angle for an increasing number of UEs. The metric $\kappa(\theta)$ has been obtained by averaging $1000$ Monte Carlo realizations of $\widehat{\HH}_{i}^{(K)}$.}
    \label{proj loss}
		\end{center}
  \vspace{-5mm}
\end{figure}

\begin{figure*}[!t]
\begin{center}
   \resizebox{0.95\textwidth}{!}{
    \includegraphics[scale=1.2]{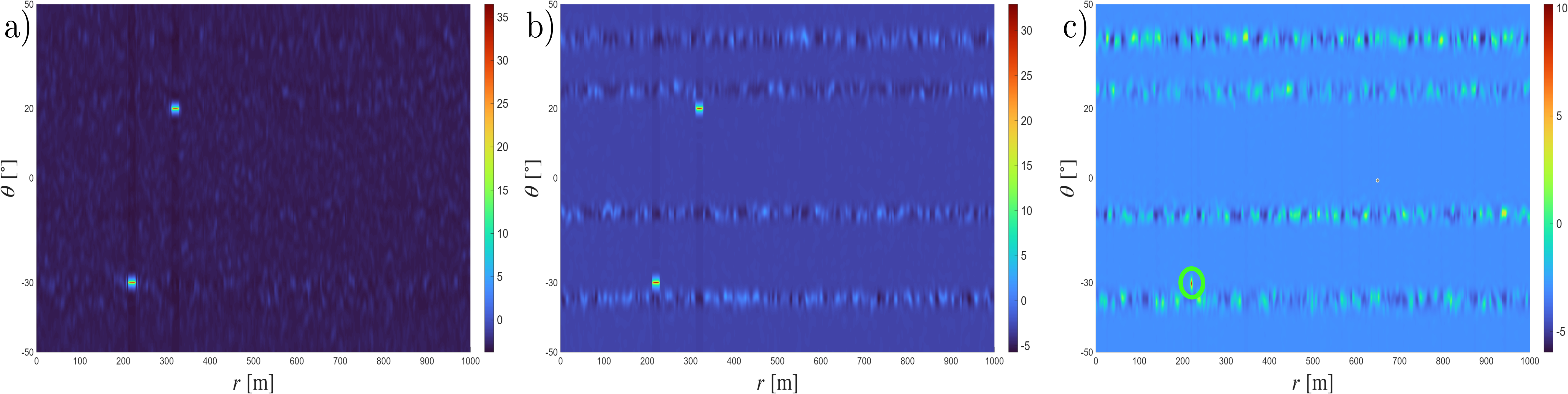}}
      \vspace{-1mm}
	 \caption{\review{Range-angle map in dB scale associated with the sensing stream, for different levels of eigenvector perturbation, namely a) $\eta_\textrm{vec}=-30$\,dB, b) $\eta_\textrm{vec}=-25$\,dB, and c) $\eta_\textrm{vec}=-20$\,dB. The same singular value subtraction and windowing of Fig. \ref{RA comm} has been applied. The \ac{MF} has been applied with perfect clutter knowledge.} }
    \label{RA sens perturbed eigvec}
		\end{center}
  \vspace{-3mm}
\end{figure*}

\begin{figure*}[!t]
\begin{center}
   \resizebox{0.95\textwidth}{!}{
    \includegraphics[scale=1.2]{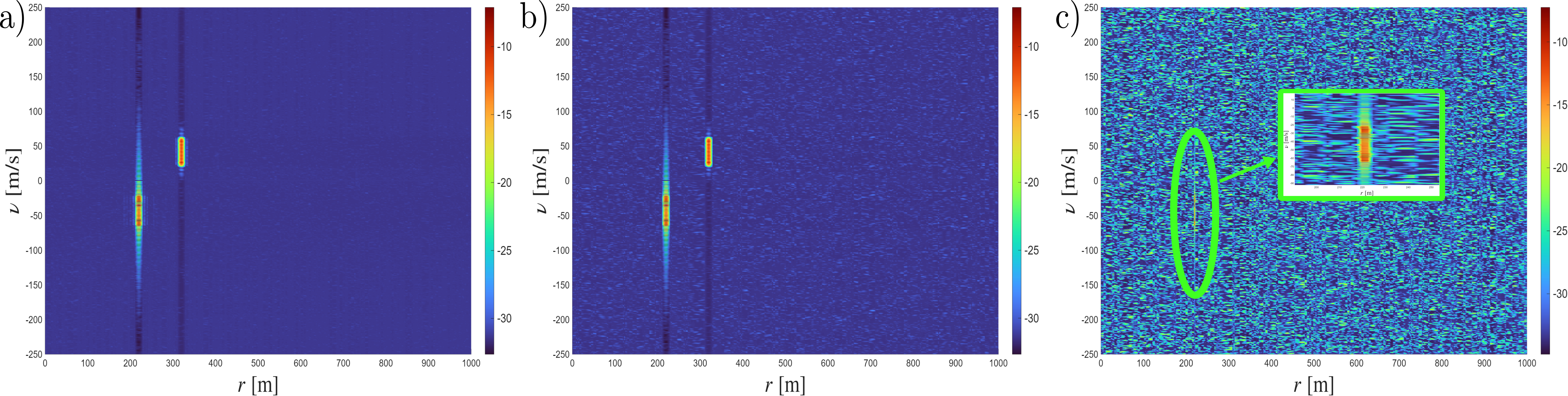}}
      \vspace{-1mm}
	 \caption{\review{Range-velocity map in dB scale associated with the sensing stream, for different levels of eigenvector perturbation, namely a) $\eta_\textrm{vec}=-30$\,dB, b) $\eta_\textrm{vec}=-25$\,dB, and c) $\eta_\textrm{vec}=-20$\,dB. } }
    \label{RV perturbed eigvec}
		\end{center}
  \vspace{-3mm}
\end{figure*}

\begin{figure}[t!]
\centering
    \begin{minipage}{\textwidth}
    \resizebox{0.48\textwidth}{!}{
%
%
\definecolor{mycolor1}{rgb}{0.06600,0.44300,0.74500}%
\definecolor{mycolor2}{rgb}{0.86600,0.32900,0.00000}%
\definecolor{mycolor3}{rgb}{0.92900,0.69400,0.12500}%
\definecolor{mycolor4}{rgb}{0.52100,0.08600,0.81900}%
\definecolor{mycolor5}{rgb}{0.23100,0.66600,0.19600}%
\definecolor{mycolor6}{rgb}{0.18400,0.74500,0.93700}%
\definecolor{mycolor7}{rgb}{0.12941,0.12941,0.12941}%
\begin{tikzpicture}

\begin{axis}[%
width=0.5\textwidth,
height=1.2in,
at={(1.888in,0.962in)},
scale only axis,
clip=false,
xmin=-29,
xmax=-15,
ymin=-0.0498360148884462,
ymax=37,
ylabel style={font=\color{mycolor7}},
ylabel={$\textrm{TFR}_S$ [dB]},
axis background/.style={fill=white},
axis x line*=bottom,
axis y line*=left,
xmajorgrids,
ymajorgrids,
legend style={legend cell align=left, align=left}
]


\addlegendimage{ mycolor1, line width=1pt}
\addlegendentry{Perfect}

\addlegendimage{ mycolor2, line width=1pt}
\addlegendentry{Estimated}

 \node[
    draw=black,
    fill=white,
    font=\small,
    anchor=north west,
    align=left,
    inner sep=4pt
] at (rel axis cs:0.8,0.62) {
\raisebox{0.3ex}{\tikz \draw[thick,solid] plot coordinates {(0,0) (0.25,0) (0.5,0)};}~$\epsilon=0.1$\\
\raisebox{0.3ex}{\tikz \draw[thick,dashed]plot coordinates {(0,0) (0.25,0) (0.5,0)};}~$\epsilon=0.5$\\
\raisebox{0.3ex}{\tikz \draw[thick,dotted]plot coordinates {(0,0) (0.25,0) (0.5,0)};}~$\epsilon=0.9$
};


\addplot [color=mycolor1, line width=1.0pt, mark=square, mark options={solid, mycolor1, mark size=2pt}, forget plot]
  table[row sep=crcr]{%
-29	38.7567627473368\\
-28	38.7119267360251\\
-27	38.6065089799078\\
-26	38.3977167137296\\
-25	37.9337816900136\\
-24	37.0096612383349\\
-23	35.3206572854011\\
-22	32.203681747651\\
-21	27.4391643793861\\
-20	21.318928416066\\
-19	11.1373720404954\\
-18	0.492233594806116\\
-17	0.087496958118806\\
-16	0.0714487832073024\\
-15	0.0113884319225525\\
};
\addlegendentry{Perfect, $\epsilon=0.1$}

\addplot [color=mycolor1, line width=1.0pt, mark=square, dashed, mark options={solid, mycolor1, mark size=2pt},forget plot]
  table[row sep=crcr]{
-29	38.5862381499098\\
-28	38.3666523724387\\
-27	37.8897875243779\\
-26	36.9452267948396\\
-25	35.3039887566894\\
-24	32.4513564013384\\
-23	27.7709072946171\\
-22	20.8198965361792\\
-21	8.93893572763962\\
-20	0.234694455908878\\
-19	0.00788281900580158\\
-18	-0.0504668463218847\\
-17	-0.0574357658892481\\
-16	-0.0475576812860274\\
-15	-0.0636180757169778\\
};
\addlegendentry{Perfect, $\epsilon=0.5$}

\addplot [color=mycolor1, line width=1.0pt, mark=square, dotted, mark options={solid, mycolor1, mark size=2pt},forget plot]
  table[row sep=crcr]{%
-29	38.4145572017513\\
-28	38.0052150630402\\
-27	37.1979506444263\\
-26	35.7570766301028\\
-25	33.373000151368\\
-24	29.3665633191602\\
-23	23.2674462381268\\
-22	13.6293125894304\\
-21	0.512562655799985\\
-20	0.036480402843195\\
-19	-0.0287841722542995\\
-18	-0.062783570824761\\
-17	-0.0645948509619449\\
-16	-0.0544781175978382\\
-15	-0.068351777866133\\
};
\addlegendentry{Perfect, $\epsilon=0.9$}

\addplot [color=mycolor2, line width=1.0pt, mark=triangle, mark options={solid, mycolor2, mark size=2pt},forget plot]
  table[row sep=crcr]{%
-29	20.0867340123476\\
-28	20.6422686187203\\
-27	19.2699537988167\\
-26	16.8228755560594\\
-25	13.6721282134273\\
-24	5.21025501881418\\
-23	0.776751377702071\\
-22	0.17338537415069\\
-21	0.014779481674243\\
-20	-0.0413456120166485\\
-19	-0.0683922243106119\\
-18	-0.0554524098690111\\
-17	-0.0758687022569013\\
-16	-0.077841706529349\\
-15	-0.0776836422703118\\
};
\addlegendentry{Estimated, $\epsilon=0.1$}

\addplot [color=mycolor2, line width=1.0pt, mark=triangle, dashed, mark options={solid, mycolor2, mark size=2pt},forget plot]
  table[row sep=crcr]{%
-29	17.3834705651757\\
-28	12.6338299571109\\
-27	3.04578732607709\\
-26	0.829251691242246\\
-25	0.0667007863872381\\
-24	0.00592050960714392\\
-23	-0.0131701754452671\\
-22	-0.0489258676471952\\
-21	-0.0694883006790377\\
-20	-0.047735959270179\\
-19	-0.0444873040775822\\
-18	-0.0442847123913284\\
-17	-0.0625326819465372\\
-16	-0.0652841455758922\\
-15	-0.0585372379097743\\
};
\addlegendentry{Estimated, $\epsilon=0.5$}

\addplot [color=mycolor2, line width=1.0pt, mark=triangle,dotted, mark options={solid, mycolor2, mark size=2pt},forget plot]
  table[row sep=crcr]{%
-29	8.4692577997147\\
-28	1.40504365153107\\
-27	0.169752097859662\\
-26	0.0772240130618925\\
-25	0.0382384203133033\\
-24	-0.0372787197343815\\
-23	-0.0608356651085957\\
-22	-0.0647245800080629\\
-21	-0.0780667319589494\\
-20	-0.0907356776646453\\
-19	-0.0725348762662874\\
-18	-0.084187908353739\\
-17	-0.057724731692716\\
-16	-0.086112784694495\\
-15	-0.0756365760285842\\
};
\addlegendentry{Estimated, $\epsilon=0.9$}

\node at (axis description cs:0.5,-0.2) {(a)};

\end{axis}
\end{tikzpicture}%
}
    \end{minipage}
\vspace{2mm}
   \begin{minipage}{\textwidth}
    \resizebox{0.48\textwidth}{!}{
%
%
\definecolor{mycolor1}{rgb}{0.06600,0.44300,0.74500}%
\definecolor{mycolor2}{rgb}{0.86600,0.32900,0.00000}%
\definecolor{mycolor3}{rgb}{0.92900,0.69400,0.12500}%
\definecolor{mycolor4}{rgb}{0.52100,0.08600,0.81900}%
\definecolor{mycolor5}{rgb}{0.23100,0.66600,0.19600}%
\definecolor{mycolor6}{rgb}{0.18400,0.74500,0.93700}%
\definecolor{mycolor7}{rgb}{0.12941,0.12941,0.12941}%
\begin{tikzpicture}

\begin{axis}[%
width=0.5\textwidth,
height=1.2in,
clip=false,
at={(1.888in,0.962in)},
scale only axis,
xmin=-29,
xmax=-15,
xlabel={$\eta_\textrm{vec}$ [dB]},
ymin=0,
ymax=35,
ylabel style={font=\color{mycolor7}},
ylabel={$\textrm{TFR}_S$ [dB]},
axis background/.style={fill=white},
axis x line*=bottom,
axis y line*=left,
xmajorgrids,
ymajorgrids,
legend style={legend cell align=left, align=left}
]


\addlegendimage{ mycolor4, line width=1pt}
\addlegendentry{Perfect}

\addlegendimage{ mycolor5, line width=1pt}
\addlegendentry{Estimated}



\addplot [color=mycolor4, line width=1.0pt, mark=square, mark options={solid, mycolor4, mark size=2pt}, forget plot]
  table[row sep=crcr]{%
-29	33.7409155049862\\
-28	33.7547437917939\\
-27	33.7734956421611\\
-26	33.7887973510773\\
-25	33.8038668018656\\
-24	33.5993998343957\\
-23	32.9028562881079\\
-22	31.4794053505045\\
-21	28.9652594740157\\
-20	25.1925095684822\\
-19	20.1976800474611\\
-18	13.7725085744209\\
-17	4.16095507018452\\
-16	0.548455140049105\\
-15	0.584523804899829\\
};
\addlegendentry{Perfect, $\epsilon=0.1$}

\addplot [color=mycolor4, line width=1.0pt, mark=square, dashed, mark options={solid, mycolor4, mark size=2pt}, forget plot]
  table[row sep=crcr]{%
-29	33.7932568145492\\
-28	33.8209165187095\\
-27	33.837442435358\\
-26	33.6460024354292\\
-25	32.9887793544765\\
-24	31.5598223625896\\
-23	28.9031757567649\\
-22	24.6449940274918\\
-21	18.6265723448443\\
-20	9.03270978485478\\
-19	0.152476044105205\\
-18	-0.0461476280284981\\
-17	-0.0694064440197119\\
-16	-0.063098517855775\\
-15	-0.0535409539357647\\
};
\addlegendentry{Perfect, $\epsilon=0.5$}

\addplot [color=mycolor4, line width=1.0pt, mark=square, dotted, mark options={solid, mycolor4, mark size=2pt}, forget plot]
  table[row sep=crcr]{%
-29	33.824903404645\\
-28	33.850001569116\\
-27	33.725418971576\\
-26	33.2197531030645\\
-25	32.0579705045772\\
-24	29.8127256493265\\
-23	25.9980666991356\\
-22	20.3169747765074\\
-21	11.8203595011183\\
-20	0.574500378920226\\
-19	-0.0417345752148658\\
-18	-0.0710083034328244\\
-17	-0.080467400982771\\
-16	-0.0731090695059805\\
-15	-0.0662540123043632\\
};
\addlegendentry{Perfect, $\epsilon=0.9$}

\addplot [color=mycolor5, line width=1.0pt, mark=triangle, mark options={solid, mycolor5, mark size=2pt}, forget plot]
  table[row sep=crcr]{%
-29	24.3141159953019\\
-28	24.4395025928391\\
-27	23.6768042138381\\
-26	22.2301581959882\\
-25	21.5871699860024\\
-24	18.2667666073414\\
-23	13.1356668182426\\
-22	4.70700964089738\\
-21	0.332494334012038\\
-20	-0.000180546740003873\\
-19	-0.0364474508835984\\
-18	-0.0496730323023759\\
-17	-0.0712913330796993\\
-16	-0.0638076641118563\\
-15	-0.0414934023441586\\
};
\addlegendentry{Estimated,, $\epsilon=0.1$}

\addplot [color=mycolor5, line width=1.0pt, mark=triangle,dashed, mark options={solid, mycolor5, mark size=2pt}, forget plot]
  table[row sep=crcr]{%
-29	23.8302007146231\\
-28	21.2678438416656\\
-27	17.5992604647315\\
-26	9.80093661879741\\
-25	3.1395583325492\\
-24	0.181528697307802\\
-23	-0.0151984799534365\\
-22	-0.0351226253173206\\
-21	-0.0761551029153879\\
-20	-0.0819838014667062\\
-19	-0.0815057944901611\\
-18	-0.0895129741729909\\
-17	-0.0857641083253904\\
-16	-0.081964916437626\\
-15	-0.0768021786781725\\
};
\addlegendentry{Estimated,, $\epsilon=0.5$}

\addplot [color=mycolor5, line width=1.0pt, mark=triangle, dotted, mark options={solid, mycolor5, mark size=2pt}, forget plot]
  table[row sep=crcr]{%
-29	17.5465088224682\\
-28	13.9624317766028\\
-27	7.6455703222715\\
-26	1.82544775661452\\
-25	0.650203179338481\\
-24	-0.0378031029211763\\
-23	-0.0680464374352455\\
-22	-0.0969384120798836\\
-21	-0.0887503398147523\\
-20	-0.085988437177373\\
-19	-0.0911325592264281\\
-18	-0.0902603095307076\\
-17	-0.0924639017802799\\
-16	-0.0642337239384326\\
-15	-0.0651882053735939\\
};
\addlegendentry{Estimated,, $\epsilon=0.9$}

\node at (axis description cs:0.5,-0.4) {(b)};

\end{axis}
\end{tikzpicture}%
}
    \end{minipage}
\vspace{-3mm}
    \caption{
    \review{\ac{TFR} of the first target (a) and second target (b) 
     as a function of the eigenvector perturbation strength $\eta$.
     The \ac{TFR} is computed on the RA maps, such as the ones shown in Fig.\ref{RA sens perturbed eigvec}. The RA maps underwent
     the same singular value subtraction of Fig. \ref{RA comm}. This figure has been obtained by averaging the TFRs of 15 Monte Carlo realizations of $\YY$.}}
\vspace{-3mm}
\label{subfig TFR}
\end{figure}


\vspace{-2mm}
\subsection{Clutter Estimation}
We now move over to evaluating the performance of the clutter covariance estimation method proposed in Section \ref{cl est sec}. Once again the performance metric is the \ac{NMSE} between the estimated covariance matrix and the actual one, which is defined as $\textrm{NMSE}_\textrm{x}= \Vert\widehat{\BB}_\textrm{x} - \BB_\textrm{x}\Vert^2_\textrm{F}/\norm{\BB_\textrm{x} }^2_\textrm{F}$. A particular mention goes to the \ac{NMSE} labeled ``Freq.sparse", which measures how well \ac{MUSIC} can capture $\overline{\BB}_\textrm{f}$, since it is defined as  $\textrm{NMSE}_\textrm{f. sparse}= \Vert\widehat{\BB}_\textrm{f} - \overline{\BB}_\textrm{f}\Vert^2_\textrm{F}/\norm{\overline{\BB}_\textrm{f} }^2_\textrm{F}$.
Our numerical results related to clutter estimation are obtained by averaging 15 independent $\YY$.
Starting from Fig. \ref{CLNMSE_vs_Bcho}, we see that a high $B_\textrm{cho}$ greatly hinders both frequency \acp{NMSE}.
The source of this error is shown in Fig. \ref{beamp_bcho}, where the beampattern of the estimated frequency covariance matrix is compared against the true one. Here we can see that the diffusive covariance component induces an exponential decay from $r=0$: when $B_\textrm{cho}$ is high, this decay becomes very sharp, so sharp that \ac{MUSIC} confuses it for a peak originated by a clutter patch.
A similar behavior can be seen in Fig.~\ref{beamp_chi}: as $\chi$ increases, the exponential decay induced by $\widetilde{\BB}_\textrm{f}$ becomes more prominent, thereby causing \ac{MUSIC} to identify false peaks around $r=0$. The frequency-\ac{NMSE} degradation induced by an increasing $\chi$ can be seen in Fig.~\ref{CLNMSE_vs_chi}. Here, we can observe that both frequency \acp{NMSE} follow the same trend, sharply rising from $-30$ dB and converging around $0$ dB. 
Interestingly, clutter with a more prominent surface roughness and multipath appears to be beneficial for estimating $\BB_\textrm{sp}$ and $\BB_\textrm{t}$, as evidenced by the decrease in their \acp{NMSE} with $\chi$.
Lastly, Fig. \ref{CLNMSE_vs_del} shows the \ac{NMSE} progression as a function of the clutter patches' delay spreads. We observe that, as the spread increases, the \ac{NMSE} associated with the frequency component increases as well. This is to be expected, as a larger spread corresponds to less orthogonal patches, in turn spoiling the functioning of \ac{MUSIC} as it relies on said orthogonality.
However, it is interesting to notice the interplay between components: 
In Fig. \ref{CLNMSE_vs_del} we can observe that an increase in delay spread seems to be beneficial for the time and space \acp{NMSE}.
A similar behavior is observed when $\varsigma\textrm{sp}^2$ increases, as the space NMSE soars while the frequency-sparse one diminishes.
It is also worth mentioning that, as $I_c$ increases, the time \ac{NMSE} sharply decreases until converging around $-30$dB.
The NMSE progressions as a function of $I_c$ and $\varsigma\textrm{sp}^2$ have not been given dedicated figures due to space constraints.
These so-called cross-contamination effects are interesting because, while it is true that the values of the parameters shown in Fig.~\ref{CLNMSE_vs_Bcho}-\ref{CLNMSE_vs_del} are set by nature, they are not immune to the observer effect of the BS. For instance, allocating more antennas to the receiving array of the BS  may decrease the observed angular spread; within this frame, the previous figures become a sort of usage manual, explaining the consequence of increasing or decreasing a clutter parameter.

\vspace{-2mm}
\subsection{Range Maps}

\reviewAfter{We now move over to assessing the quality of the  \ac{RA} and \ac{RV} generated after clutter suppression. The peak's sharpness, their magnitude, and the clutter residuals are qualitative indicators of the system's sensing performance rather than a formal equivalent to the detection and estimation metrics. }
We compare the look of the maps associated with the stream with $s=1$, which shall be referred to as the communication stream, and the ones associated with the sensing stream.
\reviewAfter{
We start with Fig. \ref{RA comm}, showing the RA maps associated with the communication streams for the standard $\FF_i^\textrm{MMSE}$ and for its QR-normalized counterpart $\FF_i^\textrm{QR}$. The top row shows that maps generated by $\mathbf{F}^{\rm MMSE}_i$ suffer from poor range localization and uneven illumination. This is caused by the fact that $\FF_i^\textrm{MMSE}$ is designed for the UE communication link and is shaped by the clustered propagation environment, without any notion of "directionality" that can be found in a beamsweeping precoder. 
The uneven illumination achieved by $\mathbf{F}^{\rm MMSE}_i$ makes the target peaks highly sensitive to the angular proximity between the set of UE communication clusters and each target. This is exemplified by the progression from Fig \ref{RA comm}a to \ref{RA comm}c, where, as the set of clusters moves, different targets are prioritized.
The bottom row shows that QR normalization improves the map quality. Nevertheless, the illumination remains dictated by the communication propagation geometry, as the peak amplitude remains proportional to its alignment with the set of communication clusters.
We can conclude that when $\mathbf{F}^{\rm MMSE}_i$ is adopted, dedicated sensing streams can be beneficial for producing informative range maps.
}
A different situation is depicted in Fig. \ref{RA sens}, which shows the RA map associated with the sensing stream in different \ac{MF} regimes. 
Fig.~\ref{RA sens}a shows that ignoring clutter leads to unusable maps. The impact of clutter awareness is evident in Fig.~\ref{RA sens}b, where we have applied MF using the estimated covariances. As a result, the map appears clean, though it still shows horizontal stripes.
When the true covariance matrix is employed, we see that the horizontal stripes become almost perfectly canceled by the \ac{SV} subtraction, and the peaks have sensibly higher power. 
\review{
The \ac{RV} map associated with Fig.\ref{RA sens}b is quantitatively similar to the one shown in Fig.\ref{RV perturbed eigvec}a, which corresponds to a lightly-perturbed clutter regime.
The \ac{RV} map associated with \ref{RA sens}b confirms the trend of  \ref{RA sens}a, where using the true covariance in the MF process gives rise to sensibly higher peaks. 
Interestingly, the first target peak is wider in the \ac{RV} maps obtained with perfect clutter knowledge.
The \ac{RV} maps associated with \ref{RA sens}a and \ref{RA sens}b have been omitted here for brevity; a representative figure can be found in Fig.\ref{RV perturbed eigvec}a.
The \ac{RV} maps associated with Fig.\ref{RA comm} and \ref{RA sens} have been omitted as they look like noise floors without any discernible trend.
Finally, the impact of $\gamma$ on the sensing streams' RA maps is shown in Fig. \ref{TFR high gamma}: we can see that the TFR of both targets decreases slightly until $\gamma=0.9$, then drops rapidly beyond that point. This behavior allows the BS to allocate the majority of the power to the communication precoder without hindering the identifiability of the targets.
\reviewAfter{These favorable map conditions are ascribed to the point-like nature of the targets and MUSIC's robustness to SNR decreases: under harsher sensing conditions, such as a higher pathloss or distributed targets, the TFR decay with $\gamma$ would be much more pronounced.}
}
\vspace{-3mm}
\reviewAfter{\subsection{Multi-UE validation of NS projection}
We investigate the impact of multiple \acp{UE} by showing their impact on the sensing-beamformer \ac{NS} projection. We consider $K\in \{1,2,3\}$ \acp{UE} having the same $\alpha_\ue$ and $\CC_\bs$, while the UE-side covariances are generated from non-overlapping $\Delta\psi_\ue$. Each UE's estimated channel is vertically stacked into $\widehat{\HH}_i^{(K)}=\alpha_\ue [\widehat{\HH}_{i,1}^\top,\dots,\widehat{\HH}_{i,K}^\top]^\top \in \mathbb{C}^{K\Mue\times \Mbs}$, from which the NS projector is defined as 
$\boldsymbol{\Pi}^{(K)}=\mathbf{I}_\Mbs - (\widehat{\HH}_i^{(K)})^\dagger\widehat{\HH}_i^{(K)}$. The projection-induced sensing loss is measured through $\kappa(\theta)=\Vert \boldsymbol{\Pi}^{(K)}\atx(\theta)\Vert^2 / \Vert \atx(\theta) \Vert^2$, shown in Fig. \ref{proj loss}. As $K$ increases, $\widehat{\HH}_i^{(K)}$ spans a larger subspace of the BS transmit space, producing a larger projection loss. This loss stays localized around the same angular region as the UEs share the same $\CC_\bs$.
}
\subsection{Non-separable perturbation robustness}
\reviewAfter{We now provide some numerical results to characterize the robustness of the proposed clutter estimation and suppression procedure to non-separable clutter, according to the perturbation model provided in \ref{rob sec}.}
\reviewAfter{It should be noted that results associated with eigenvalue perturbation are not reported, as the generated maps resulted in being much less sensitive to them than to eigenvector perturbation.} This is ascribed to MUSIC's robustness to SNR changes and high sensitivity to subspace misalignment. 
Therefore, unless otherwise specified, we assume $\epsilon=0.5$ and $\eta_\textrm{val}=0$.
Starting from Fig.\ref{RA sens perturbed eigvec}, we see that when $\eta_\textrm{vec}=-30$\,dB the whitening procedure is still effective as target peaks are still clearly distinguishable from the noise floor; however, as $\eta_\textrm{vec}=-30$ increases, the perturbation rapidly decreases the quality of the maps, until eventually only one target survives when $\eta_\textrm{vec}=-20$\,dB.
The \ac{RV} maps depicted in Fig.\ref{RV perturbed eigvec} confirm this trend, with Fig.\ref{RV perturbed eigvec}a showing a map with distinguishable target peaks and Fig.\ref{RV perturbed eigvec}b resembling a noise floor, except with a very narrow slit in correspondence of the closest target.
Lastly, a quantitative assessment of the target's identifiability comes from  Fig. \ref{subfig TFR}: here we see the TFR of both targets on the sensing stream map as a function of the eigenvector perturbation magnitude $\eta_\textrm{vec}$ for different levels of perturbation.
As expected, all of the curves present a steady decrease as $\eta_\textrm{vec}$ increases, as the Kronecker-separability assumptions become progressively weaker. On the other hand, under small perturbation regimes (i.e., $\eta_\textrm{vec} \leq -25$\,dB and $\epsilon<0.5$), \reviewAfter{ the proposed clutter estimation and suppression procedure still produces identifiable target peaks.}

\vspace{-4mm}
\section{Conclusions}
This paper examined a \ac{MIMO} \ac{OFDM} \ac{ISAC} network that serves a \ac{UE} and performs monostatic sensing. The temporal evolution of the propagation environment induces channel aging, motivating the development of an aging-aware channel estimator and Doppler-affected clutter model. 
\reviewAfter{This clutter model allowed the development of a novel clutter estimation and suppression procedure, allowing for the generation of high-quality range maps in a clutter-dominated environment.}
Its low complexity makes it possible to be deployed on \ac{ISAC}-capable \acp{BS} operating in the presence of urban clutter. 
Our results show that, in low-to-moderate mobility regimes, reliance on a block-fading model leads to substantial performance degradation.
\reviewAfter{We have shown that under reasonable assumptions, structured clutter can be effectively modeled and mitigated, enabling the generation of good range maps. Furthermore, the proposed robustness analysis has shown that the proposed method can withstand moderate non-separability while still preserving target identifiability.}
As part of future work, we plan on extending this model to multi-static \ac{ISAC} scenarios, where perfect knowledge of the transmitted symbols is not always guaranteed.
\reviewAfter{A systematic investigation of the coupled effects introduced by the presence of multiple UEs is beyond the scope of the present paper and is left for future work.}

\bibliographystyle{IEEEtran}
\bibliography{IEEEabrv,auxiliary/biblio,JCASv10}

\begin{thebibliography}{10}
\providecommand{\url}[1]{#1}
\csname url@samestyle\endcsname
\providecommand{\newblock}{\relax}
\providecommand{\bibinfo}[2]{#2}
\providecommand{\BIBentrySTDinterwordspacing}{\spaceskip=0pt\relax}
\providecommand{\BIBentryALTinterwordstretchfactor}{4}
\providecommand{\BIBentryALTinterwordspacing}{\spaceskip=\fontdimen2\font plus
\BIBentryALTinterwordstretchfactor\fontdimen3\font minus \fontdimen4\font\relax}
\providecommand{\BIBforeignlanguage}[2]{{%
\expandafter\ifx\csname l@#1\endcsname\relax
\typeout{** WARNING: IEEEtran.bst: No hyphenation pattern has been}%
\typeout{** loaded for the language `#1'. Using the pattern for}%
\typeout{** the default language instead.}%
\else
\language=\csname l@#1\endcsname
\fi
#2}}
\providecommand{\BIBdecl}{\relax}
\BIBdecl

\bibitem{liu2022integrated}
F.~Liu, Y.~Cui, C.~Masouros, J.~Xu, T.~X. Han, Y.~C. Eldar, and S.~Buzzi, ``Integrated sensing and communications: Towards dual-functional wireless networks for {6G} and beyond,'' \emph{IEEE journal on selected areas in communications}, 2022.

\bibitem{Babu:24}
N.~Babu, C.~Masouros, C.~B. Papadias, and Y.~C. Eldar, ``Precoding for multi-cell {ISAC}: From coordinated beamforming to coordinated multipoint and bi-static sensing,'' \emph{IEEE Transactions on Wireless Communications}, vol.~23, no.~10, pp. 14\,637--14\,651, 2024.

\bibitem{keskin2022optimal}
M.~F. Keskin, F.~Jiang, F.~Munier, G.~Seco-Granados, and H.~Wymeersch, ``Optimal spatial signal design for mmwave positioning under imperfect synchronization,'' \emph{IEEE Transactions on Vehicular Technology}, vol.~71, no.~5, pp. 5558--5563, 2022.

\bibitem{Fang:23}
X.~Fang, W.~Feng, Y.~Chen, N.~Ge, and Y.~Zhang, ``Joint communication and sensing toward {6G}: Models and potential of using {MIMO},'' \emph{IEEE Internet of Things Journal}, vol.~10, no.~5, pp. 4093--4116, 2023.

\bibitem{rivetti2024secure}
S.~Rivetti, E.~Bj{\"o}rnson, and M.~Skoglund, ``Secure spatial signal design for isac in a cell-free mimo network,'' in \emph{2024 IEEE Wireless Communications and Networking Conference (WCNC)}.\hskip 1em plus 0.5em minus 0.4em\relax IEEE, 2024, pp. 01--06.

\bibitem{AnLiu:22}
A.~Liu, Z.~Huang, M.~Li, Y.~Wan, W.~Li, T.~X. Han, C.~Liu, R.~Du, D.~K.~P. Tan, J.~Lu, Y.~Shen, F.~Colone, and K.~Chetty, ``A survey on fundamental limits of integrated sensing and communication,'' \emph{IEEE Communications Surveys \& Tutorials}, vol.~24, no.~2, pp. 994--1034, 2022.

\bibitem{Liu:25}
F.~Liu, Y.-F. Liu, Y.~Cui, C.~Masouros, J.~Xu, T.~X. Han, S.~Buzzi, Y.~C. Eldar, and S.~Jin, ``Sensing with communication signals: From information theory to signal processing,'' \emph{IEEE Journal on Selected Areas in Communications}, pp. 1--1, 2025.

\bibitem{Baig:23}
M.~U. Baig, J.~Vinogradova, G.~Fodor, and C.~Moll\'en, ``Joint communication and sensing beamforming for passive object localization,'' in \emph{WSA and SCC 2023; 26th Int. {ITG} Workshop on Smart Antennas and 13th Conf. on Systems, Communications, and Coding}, 2023, pp. 1--6.

\bibitem{rivetti2024clutter}
S.~Rivetti, O.~T. Demir, E.~Bjornson, and M.~Skoglund, ``Clutter-aware target detection for {ISAC} in a millimeter-wave cell-free massive {MIMO} system,'' \emph{arXiv preprint arXiv:2411.08759}, 2024.

\bibitem{salman2024sensing}
M.~B. Salman, {\"O}.~T. Demir, and E.~Bj{\"o}rnson, ``When are sensing symbols required for {ISAC}?'' \emph{IEEE Transactions on Vehicular Technology}, vol.~73, no.~10, pp. 15\,709--15\,714, 2024.

\bibitem{Truong:13}
K.~T. {Truong} and R.~W. {Heath}, ``Effects of channel aging in massive {MIMO} systems,'' \emph{Journal of Communications and Networks}, vol.~15, no.~4, pp. 338--351, 2013.

\bibitem{Fodor:21}
G.~Fodor, J.~Vinogradova, P.~Hammarberg, K.~K. Nagalapur, Z.~T. Qi, H.~Do, R.~Blasco, and M.~U. Baig, ``{5G} new radio for automotive, rail, and air transport,'' \emph{IEEE Communications Magazine}, vol.~59, no.~7, pp. 22--28, 2021.

\bibitem{kong2015sum}
C.~Kong, C.~Zhong, A.~K. Papazafeiropoulos, M.~Matthaiou, and Z.~Zhang, ``Sum-rate and power scaling of massive {MIMO} systems with channel aging,'' \emph{IEEE transactions on communications}, vol.~63, no.~12, pp. 4879--4893, 2015.

\bibitem{fodor2023optimizing}
S.~Fodor, G.~Fodor, D.~G{\"u}rg{\"u}no{\u{g}}lu, and M.~Telek, ``Optimizing pilot spacing in {MU-MIMO} systems operating over aging channels,'' \emph{IEEE Transactions on Communications}, vol.~71, no.~6, pp. 3708--3720, 2023.

\bibitem{Daei:25}
S.~Daei, G.~Fodor, M.~Skoglund, and M.~Telek, ``Toward optimal pilot spacing and power control in multi-antenna systems operating over non-stationary {Rician} aging channels,'' \emph{IEEE Transactions on Communications}, vol.~73, no.~6, pp. 3761--3777, 2025.

\bibitem{Bjornson2015b}
E.~Bj{\"{o}}rnson, M.~Matthaiou, and M.~Debbah, ``Massive {MIMO} with non-ideal arbitrary arrays: Hardware scaling laws and circuit-aware design,'' \emph{IEEE Transactions on Wireless Communications}, vol.~14, no.~8, pp. 4353--4368, 2015.

\bibitem{Yuan:20}
J.~{Yuan}, H.~Q. {Ngo}, and M.~{Matthaiou}, ``Machine learning-based channel prediction in massive {MIMO} with channel aging,'' \emph{IEEE Transactions on Wireless Communications}, vol.~19, no.~5, pp. 2960--2973, 2020.

\bibitem{chen2023ripoff}
J.~Chen, X.~Wang, and Y.-C. Liang, ``Impact of channel aging on dual-function radar-communication systems: Performance analysis and resource allocation,'' \emph{IEEE Transactions on Communications}, vol.~71, no.~8, pp. 4972--4987, 2023.

\bibitem{luo2024yolo}
H.~Luo, F.~Gao, H.~Lin, S.~Ma, and H.~V. Poor, ``{YOLO}: An efficient {Terahertz} band integrated sensing and communications scheme with beam squint,'' \emph{IEEE Transactions on Wireless Communications}, vol.~23, no.~8, pp. 9389--9403, 2024.

\bibitem{lu2025mimoclutt}
Q.~Lu, K.~Yang, Z.~Zhang, and T.-K. Truong, ``{MIMO OFDM} robust transceiver design for clutter suppression in {ISAC} systems,'' \emph{IEEE Transactions on Vehicular Technology}, 2025.

\bibitem{Vinogradova:23}
J.~Vinogradova and G.~Fodor, ``On target detection in the presence of clutter in joint communication and sensing cellular networks,'' in \emph{2023 16th International Conference on Signal Processing and Communication System {(ICSPCS)}}, 2023, pp. 01--10.

\bibitem{demir2024ris}
{\"O}.~T. Demir and E.~Bj{\"o}rnson, ``{RIS}-assisted {ISAC}: Precoding and phase-shift optimization for mono-static target detection,'' \emph{arXiv preprint arXiv:2410.06855}, 2024.

\bibitem{greenewald2016robust}
K.~Greenewald, E.~Zelnio, and A.~H. Hero, ``Robust {SAR STAP} via {Kronecker} decomposition,'' \emph{IEEE Transactions on Aerospace and Electronic Systems}, vol.~52, no.~6, pp. 2612--2625, 2016.

\bibitem{ward1998space}
J.~Ward, ``Space-time adaptive processing for airborne radar,'' in \emph{IEE Colloquium on Space-Time Adaptive Processing}.\hskip 1em plus 0.5em minus 0.4em\relax IET, 1998, pp. 2--1.

\bibitem{marzetta2016fundamentals}
T.~L. Marzetta and H.~Yang, \emph{Fundamentals of massive {MIMO}}.\hskip 1em plus 0.5em minus 0.4em\relax Cambridge University Press, 2016.

\bibitem{daei2024improved}
S.~Daei, M.~Skoglund, and G.~Fodor, ``Improved downlink channel estimation in time-varying {FDD} massive {MIMO} systems,'' in \emph{2024 IEEE 25th International Workshop on Signal Processing Advances in Wireless Communications (SPAWC)}.\hskip 1em plus 0.5em minus 0.4em\relax IEEE, 2024, pp. 571--575.

\bibitem{liyanaarachchi2023joint}
S.~D. Liyanaarachchi, T.~Riihonen, C.~B. Barneto, and M.~Valkama, ``Joint mimo communications and sensing with hybrid beamforming architecture and {OFDM} waveform optimization,'' \emph{IEEE Transactions on Wireless Communications}, 2023.

\bibitem{bengtsson2006some}
M.~Bengtsson and P.~Zetterberg, ``Some notes on the {Kronecker} model,'' \emph{EURASIP Journal on Wireless Communications and Networking}, 2006.

\bibitem{debbah2022uplink}
Z.~Wang, J.~Zhang, B.~Ai, C.~Yuen, and M.~Debbah, ``Uplink performance of cell-free massive {MIMO} with multi-antenna users over jointly-correlated {Rayleigh} fading channels,'' \emph{IEEE Transactions on Wireless Communications}, vol.~21, no.~9, pp. 7391--7406, 2022.

\bibitem{38901}
3GPP, ``Study on channel model for frequencies from 0.5 to 100 {GHz},'' 3GPP, Technical Specification (TS) {38.901}, 09 2025, version 19.1.0.

\bibitem{truong2013effects}
K.~T. Truong and R.~W. Heath, ``Effects of channel aging in massive {MIMO} systems,'' \emph{Journal of Communications and Networks}, vol.~15, no.~4, pp. 338--351, 2013.

\bibitem{abeida2010data}
H.~Abeida, ``Data-aided snr estimation in time-variant {Rayleigh} fading channels,'' \emph{IEEE transactions on signal processing}, vol.~58, no.~11, pp. 5496--5507, 2010.

\bibitem{brunner2024bistatic}
D.~Brunner, L.~G. de~Oliveira, C.~Muth, S.~Mandelli, M.~Henninger, A.~Diewald, Y.~Li, M.~B. Alabd, L.~Schmalen, T.~Zwick \emph{et~al.}, ``Bistatic {OFDM-based ISAC} with over-the-air synchronization: System concept and performance analysis,'' \emph{IEEE Transactions on Microwave Theory and Techniques}, vol.~73, no.~5, pp. 3016--3029, 2024.

\bibitem{NextGAlliance2025}
\BIBentryALTinterwordspacing
{Next G Alliance}, ``Channel measurements and modeling for joint/integrated communication and sensing, as well as 7--24 ghz communication,'' ATIS, Tech. Rep., 2025. [Online]. Available: \url{https://nextgalliance.org/white_papers/channel-measurements-and-modeling-for-joint-integrated-communication-and-sensing-as-well-as-7-24-ghz-communication/}
\BIBentrySTDinterwordspacing

\bibitem{skolnik2008radar}
M.~I. Skolnik, \emph{Radar handbook}.\hskip 1em plus 0.5em minus 0.4em\relax McGraw-Hill, 2008.

\bibitem{duan2024analytical}
T.~Duan, P.~Shui, J.~Wang, and S.~Xu, ``Analytical coherent detection in high-resolution dual-polarimetric sea clutter with independent inverse gamma textures,'' \emph{Remote Sensing}, vol.~16, no.~8, p. 1315, 2024.

\bibitem{demir2022RISP}
{\"O}.~T. Demir and E.~Bj{\"o}rnson, ``Is channel estimation necessary to select phase-shifts for {RIS}-assisted massive {MIMO}?'' \emph{IEEE Transactions on Wireless Communications}, vol.~21, no.~11, pp. 9537--9552, 2022.

\bibitem{abdi2002space}
A.~Abdi and M.~Kaveh, ``A space-time correlation model for multielement antenna systems in mobile fading channels,'' \emph{IEEE Journal on Selected Areas in communications}, vol.~20, no.~3, pp. 550--560, 2002.

\bibitem{demir2022channel}
{\"O}.~T. Demir, E.~Bj{\"o}rnson, and L.~Sanguinetti, ``Channel modeling and channel estimation for holographic massive {MIMO} with planar arrays,'' \emph{IEEE Wireless Communications Letters}, vol.~11, no.~5, pp. 997--1001, 2022.

\bibitem{wang2003efficient}
C.-X. Wang and M.~Patzold, ``Efficient simulation of multiple cross-correlated {Rayleigh} fading channels,'' in \emph{14th IEEE Proceedings on Personal, Indoor and Mobile Radio Communications, 2003. PIMRC 2003.}, vol.~2.\hskip 1em plus 0.5em minus 0.4em\relax IEEE, 2003, pp. 1526--1530.

\bibitem{greenewald2013kronecker}
K.~Greenewald, T.~Tsiligkaridis, and A.~O. Hero, ``Kronecker sum decompositions of space-time data,'' in \emph{2013 5th IEEE International Workshop on Computational Advances in Multi-Sensor Adaptive Processing (CAMSAP)}.\hskip 1em plus 0.5em minus 0.4em\relax IEEE, 2013, pp. 65--68.

\bibitem{Werner:08}
K.~Werner, M.~Jansson, and P.~Stoica, ``On estimation of covariance matrices with {Kronecker} product structure,'' \emph{IEEE Transactions on Signal Processing}, vol.~56, no.~2, pp. 478--491, 2008.

\bibitem{henninger2023crap}
M.~Henninger, S.~Mandelli, A.~Grudnitsky, T.~Wild, and S.~ten Brink, ``{CRAP}: Clutter removal with acquisitions under phase noise,'' in \emph{2023 2nd International Conference on 6G Networking (6GNet)}.\hskip 1em plus 0.5em minus 0.4em\relax IEEE, 2023, pp. 1--8.

\bibitem{henninger2024crap2}
M.~Henninger, S.~Mandelli, A.~Grudnitsky, and S.~ten Brink, ``{CRAP} part ii: Clutter removal with continuous acquisitions under phase noise,'' in \emph{2024 Joint European Conference on Networks and Communications \& 6G Summit (EuCNC/6G Summit)}.\hskip 1em plus 0.5em minus 0.4em\relax IEEE, 2024, pp. 416--421.

\bibitem{sherman2023eigenstructure}
P.~J. Sherman, ``On the eigenstructure of the ar (1) covariance,'' in \emph{2023 IEEE Statistical Signal Processing Workshop (SSP)}.\hskip 1em plus 0.5em minus 0.4em\relax IEEE, 2023, pp. 6--10.

\bibitem{kama2024downlink}
E.~B. Kama, J.~Kim, and E.~Bj{\"o}rnson, ``Downlink pilots are essential for cell-free massive {MIMO} with multi-antenna users,'' in \emph{2024 IEEE Wireless Communications and Networking Conference (WCNC)}.\hskip 1em plus 0.5em minus 0.4em\relax IEEE, 2024, pp. 1--6.

\bibitem{he2024mse}
Z.~He, H.~Shen, W.~Xu, Y.~C. Eldar, and X.~You, ``{MSE}-based training and transmission optimization for {MIMO ISAC} systems,'' \emph{IEEE Transactions on Signal Processing}, 2024.

\bibitem{kay2013fundamentals}
S.~M. Kay, \emph{Fundamentals of statistical signal processing: Practical algorithm development}.\hskip 1em plus 0.5em minus 0.4em\relax Pearson Education, 2013, vol.~3.

\bibitem{fodor2021performance}
G.~Fodor, S.~Fodor, and M.~Telek, ``Performance analysis of a linear {MMSE} receiver in time-variant {Rayleigh} fading channels,'' \emph{IEEE Transactions on Communications}, vol.~69, no.~6, pp. 4098--4112, 2021.

\bibitem{stoica2005spectral}
P.~Stoica, R.~L. Moses \emph{et~al.}, \emph{Spectral analysis of signals}.\hskip 1em plus 0.5em minus 0.4em\relax Pearson Prentice Hall Upper Saddle River, NJ, 2005, vol. 452.

\bibitem{rohling2007radar}
H.~Rohling, ``Radar {CFAR} thresholding in clutter and multiple target situations,'' \emph{IEEE transactions on aerospace and electronic systems}, no.~4, pp. 608--621, 2007.

\bibitem{zhou2011godec}
T.~Zhou and D.~Tao, ``Godec: Randomized low-rank \& sparse matrix decomposition in noisy case,'' in \emph{Proceedings of the 28th International Conference on Machine Learning, ICML 2011}, 2011.

\bibitem{richards2010principles}
M.~A. Richards, J.~A. Scheer, and W.~A. Holm, \emph{Principles of modern radar: basic principles}.\hskip 1em plus 0.5em minus 0.4em\relax IET, 2010.

\bibitem{stoica2002music}
P.~Stoica and A.~Nehorai, ``{MUSIC}, maximum likelihood, and {Cram\'er-Rao} bound,'' \emph{IEEE Transactions on Acoustics, speech, and signal processing}, vol.~37, no.~5, pp. 720--741, 2002.

\bibitem{izquierdo2000signal}
M.~Izquierdo, M.~Hernandez, O.~Graullera, and J.~Anaya, ``Signal-to-noise ratio enhancement based on the whitening transformation of colored structural noise,'' \emph{Ultrasonics}, vol.~38, no. 1-8, pp. 500--502, 2000.

\bibitem{sun2017kronecker}
G.~Sun, Z.~He, F.~Jia, and R.~Li, ``{Kronecker} product {PCA} for structured covariance matrix of airborne radar {STAP},'' in \emph{2017 IEEE Radar Conference (RadarConf)}.\hskip 1em plus 0.5em minus 0.4em\relax IEEE, 2017, pp. 1015--1019.

\bibitem{wajid2009robust}
I.~Wajid, Y.~C. Eldar, and A.~Gershman, ``Robust downlink beamforming using covariance channel state information,'' in \emph{2009 IEEE International Conference on Acoustics, Speech and Signal Processing}.\hskip 1em plus 0.5em minus 0.4em\relax IEEE, 2009, pp. 2285--2288.

\bibitem{demir2021foundations}
{\"O}.~T. Demir, E.~Bj{\"o}rnson, L.~Sanguinetti \emph{et~al.}, ``Foundations of user-centric cell-free massive {MIMO},'' \emph{Foundations and Trends{\textregistered} in Signal Processing}, vol.~14, no. 3-4, pp. 162--472, 2021.

\bibitem{kim2019improvement}
T.-H. Kim, H.-W. Jeon, J.-H. Shin, and Y.-D. Kang, ``Improvement of detection ranges for targets in sidelobe clutter surroundings by sigma-delta stap for airborne radars,'' \emph{Journal of Electromagnetic Engineering and Science}, vol.~19, no.~4, pp. 234--238, 2019.

\end{thebibliography}
\end{document}